\documentclass[twocolumn]{aastex631}
\usepackage{graphicx}
\usepackage[varg]{txfonts}
\newcommand{\lta}{\mathrel{\hbox{\raise 0.6 ex \hbox{$<$}\kern
                   -1.8 ex\lower .5 ex\hbox{$\sim$}}}}
\newcommand{\gta}{\mathrel{\hbox{\raise 0.6 ex \hbox{$>$}\kern
                   -1.7 ex\lower .5 ex\hbox{$\sim$}}}}
\DeclareMathAlphabet{\mathsc}{T1}{cmr}{m}{sc}

\begin{document}

\title[Zeeman Doppler maps. II: eschewing physics]{Zeeman Doppler maps. II:
  the perils of eschewing physics}

\author[0000-0003-3004-4701]{M.J.~Stift}
\affiliation{Kuffner-Sternwarte, Johann~Staud-Strasse 10, A-1160~Wien, Austria}
\affiliation{Universit{\"a}t~Wien, Universit{\"a}tsring~1, 1010~Wien}
\author[0000-0002-2543-3836]{F.~Leone}
\affiliation {Dipartimento di Fisica e Astronomia, Universit{\`a} di Catania,
  Sezione Astrofisica, Via S. Sofia 78, I-95123 Catania, Italy}
\correspondingauthor{M.J.~Stift}
\email{stift@ada2012.eu}

\received{2021}

\label{firstpage}

\begin{abstract}
For the observational modeling of horizontal abundance distributions and of magnetic
geometries in chemically peculiar (CP) stars, Zeeman Doppler mapping (ZDM) has become the
method of choice. Comparisons between abundance maps obtained for CP stars and predictions
from numerical simulations of atomic diffusion have always proved unsatisfactory, with the
blame routinely put on theory. Expanding a previous study aimed at clarifying the question
of the uniqueness of ZDM maps, this  paper inverts the roles between observational
modeling and time-dependent diffusion results, casting a cold eye on essential assumptions
and algorithms underlying ZDM, in particular the Tikhonov-style regularization
functionals, from 1D to 3D. We show that these have been established solely for
mathematical convenience, but that they in no way reflect the physical reality in the
atmospheres of magnetic CP stars. Recognizing that the observed strong magnetic fields in
most well-mapped stars require the field geometry to be force-free, we demonstrate that
many published maps do not meet this condition. There follows a discussion of the frequent
changes in magnetic and abundance maps of well observed stars and a caveat concerning the
use of least squares deconvolution in ZDM analyses. It emerges that because of the
complexity and non-linearity of the field-dependent chemical stratifications, Tikhonov
based ZDM inversions cannot recover the true abundance and magnetic geometries. As our
findings additionally show, there is no way to define a physically meaningful 3D
regularization functional instead. ZDM remains dysfunctional and does not provide any
observational constraints for the modeling of atomic diffusion.
\end{abstract}


  
\section{Introduction}
\label{intro}

Over the past decades, (Zeeman) Doppler mapping (ZDM) has established itself as the most
popular method for the analysis of abundances and magnetic fields in upper main sequence
chemically peculiar (CP) stars. Inverting the observed intensity profiles of spectral
lines, one can in principle map elemental abundances over the stellar disk; by adding
full Stokes profiles, the reconstruction of the stellar vector magnetic field also
becomes possible. Unfortunately, direct inversion is impossible, since mathematically
we are faced with an ill-posed problem offering a huge variety of possible solutions
all of which are able to reproduce the observed profiles to the same level of accuracy.
There is thus the need for a constraint which leads to a unique solution for the
magnetic field geometry or the maps of the different chemical elements. Ideally, such
a constraint should reflect the physics of the abundances encountered in CP stars and/or
of the magnetic fields that are the cause of patches, spots or rings, but in real life
that has so far never been the case. It has proved far more expedient to resort to
the use of mathematically convenient penalty/regularization functions which are easy
to implement. Historically, Doppler mapping started with maximum entropy regularization
(see \citealt{VogtVoPeHa1987}) but later Tikhonov regularization has taken over as far
as CP stars are concerned (see e.g. \citealt{Piskunov2001}). While the maximum entropy
image has the least amount of spatial information, the use of the Tikhonov functional
leads to the smoothest possible solution. Still, provided the inversion is based on high
S/N ratio spectra in all 4 Stokes parameters, well sampled over the rotational phases,
\citet{Piskunov2001} claimed that ``the exact form of the regularization function is not
important'' and ``although it was never proven that the DI or MDI problem has a unique
solution if a perfect data set is available, extensive experiments carried by many
authors suggest that it is the case''. This assumedly unique solution would represent
the ``true'' magnetic and abundance maps.

Over time, the initial assumptions under which Doppler mapping yielded meaningful
results seem to have been largely forgotten. The famous ``Vogtstar'' used by
\citet{VogtVoPeHa1987} presented perfectly black letters written onto a bright stellar
surface; the ``seven-spot'' image was devised exactly in the same vein. 15 years later,
\citet{KochukhovKoPi2002} (=KP02) resorted to  similar test-cases with 3 well distributed
high-contrast (1.5\,dex) stellar spots. In both papers, the spots were simply taken as
geometrical artifacts, in no way resulting from any atmospheric process or physically
related to the magnetic field geometry, implying that a purely mathematical approach to
the regularization function is absolutely appropriate. What the tests demonstrate is that
under such extremely restrictive and artificial assumptions, brightness or abundance
mapping gives results that are in tolerable agreement with the input data, even though
true abundances might be missed by 0.5\,dex and more (see Figure\,3 of KP02 but also
\citet{Kochukhov2017}). The convergence of a ZDM code towards a solution similar to
such a oversimplified input map merely means that basically the algorithm works correctly,
but by no means do these tests validate the KP02 statement: ``we believe that the code
can be successfully applied to the imaging of global stellar magnetic fields and abundance
distributions of an arbitrary complexity''.

While the decisive role of magnetic fields in the outer solar layers is well recognized
and nobody would put into doubt that the movements of charged particles react most
sensitively to the direction of the magnetic field, none of this insight has ever
entered the ``standard'' ZDM procedure as applied to CP stars and based in its entirety
on one single proprietary code. Inversions based on physics-free regularization
functionals have been taken as unassailable pillars of astrophysical knowledge,
putting the blame for theoretical predictions which are at variance with ZDM maps
squarely and entirely on the alleged lack of sophistication of diffusion theory and
the numerical codes (e.g. \citealt{NesvacilNeLuShetal2012}).

What about the possibility that the discrepancies between theory and observations
are due to the assumptions underlying the ``standard'' ZDM inversion procedure?
Could it be that mathematical expediency has triumphed over physics? Tikhonov
regularization in various shades of simplicity has been applied to everything,
from straightforward abundance maps to 1D (i.e. globally constant) stratifications
and to 3D stratifications, but do magnetic stars really conform to this incredibly
plain picture? In the next section a review of the latest developments in the
modeling of atomic diffusion strongly suggests that neither Tikhonov nor maximum
entropy functionals provide the means to recover true horizontal abundance
inhomogeneities, even less so 3D abundance stratifications, in accord with 
previous findings of \citet{StiftLeone2017a}. In the subsequent section,
a critical look at the formulae governing the inversion of the magnetic field reveals
that the force-free condition of the magnetic field geometry in a number of strongly
magnetic CP stars has been neglected/overlooked, resulting in spurious magnetic maps
and ensuing abundance maps that are necessarily spurious, too. A short discussion of
the 3D nature chemical stratifications similar to what has already been sketched by
\citet{StiftLeone2017a} leads to the conclusion that a physically meaningful
regularization functional cannot be devised, leaving ZDM open to convergence towards
solutions at variance with firmly established astrophysical knowledge.

\begin{figure*}
\begin{centering}
\includegraphics[width=50mm, height=57mm, angle=270]{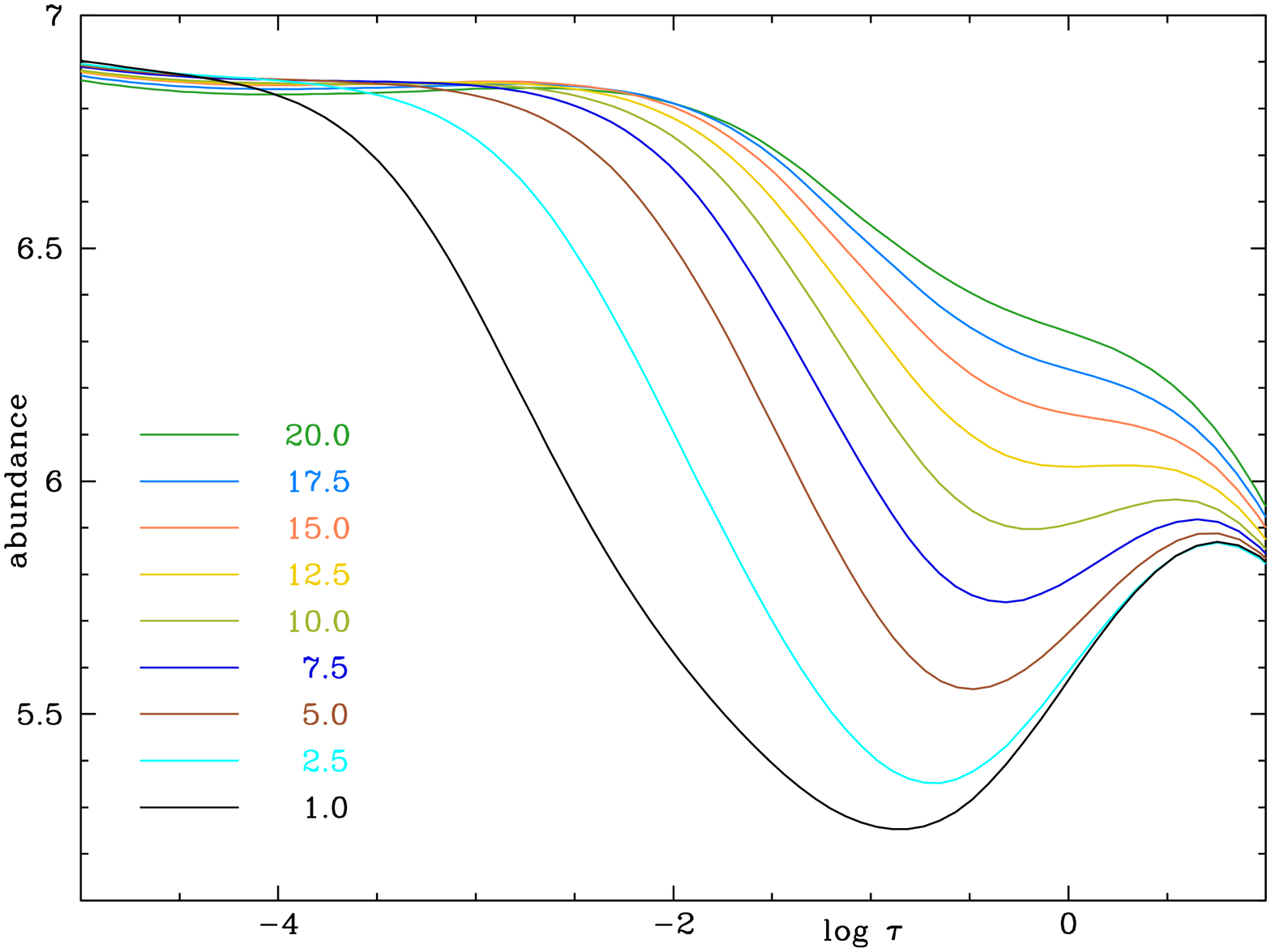}
\includegraphics[width=50mm, height=57mm, angle=270]{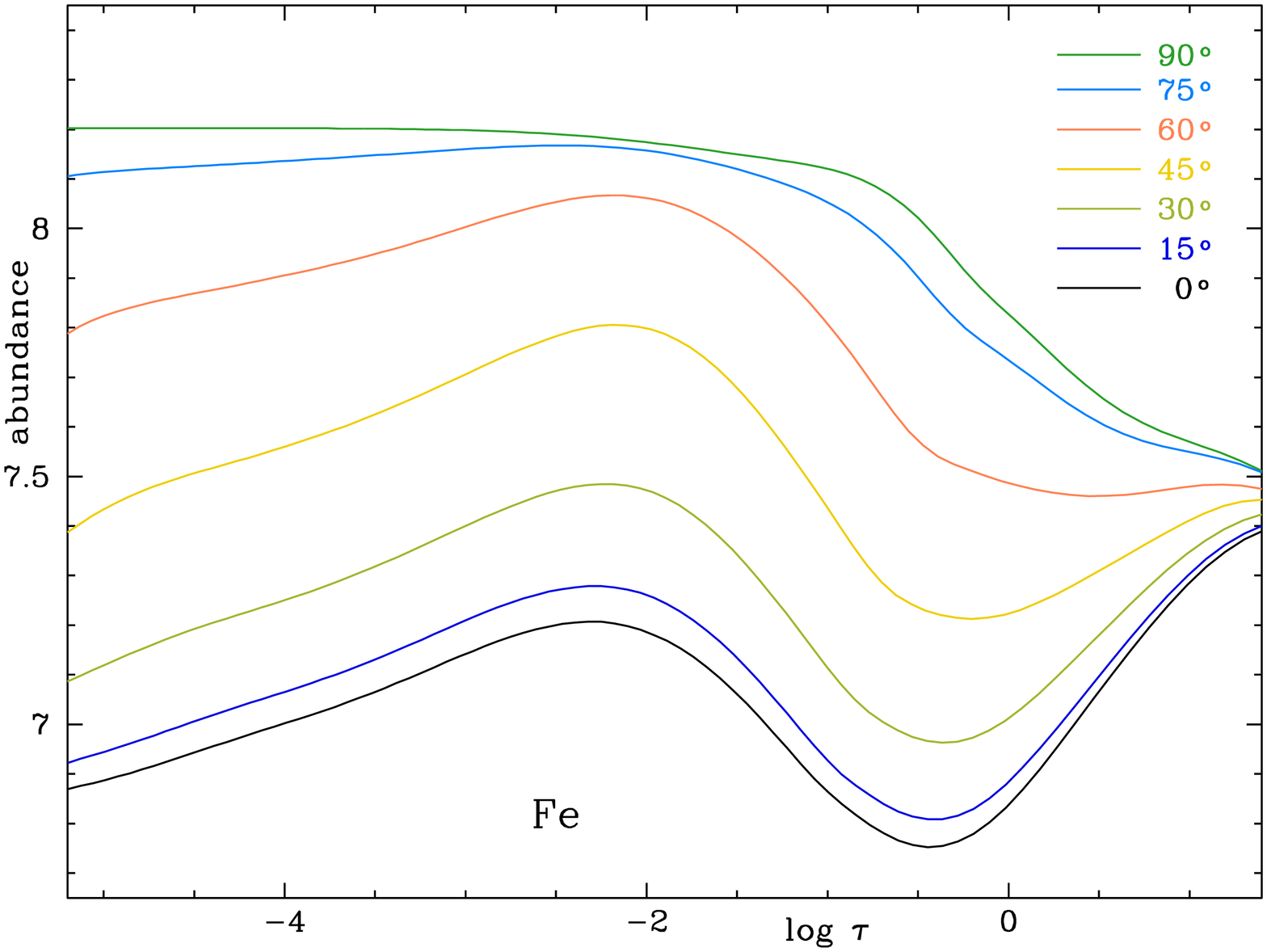}
\includegraphics[width=50mm, height=57mm, angle=270]{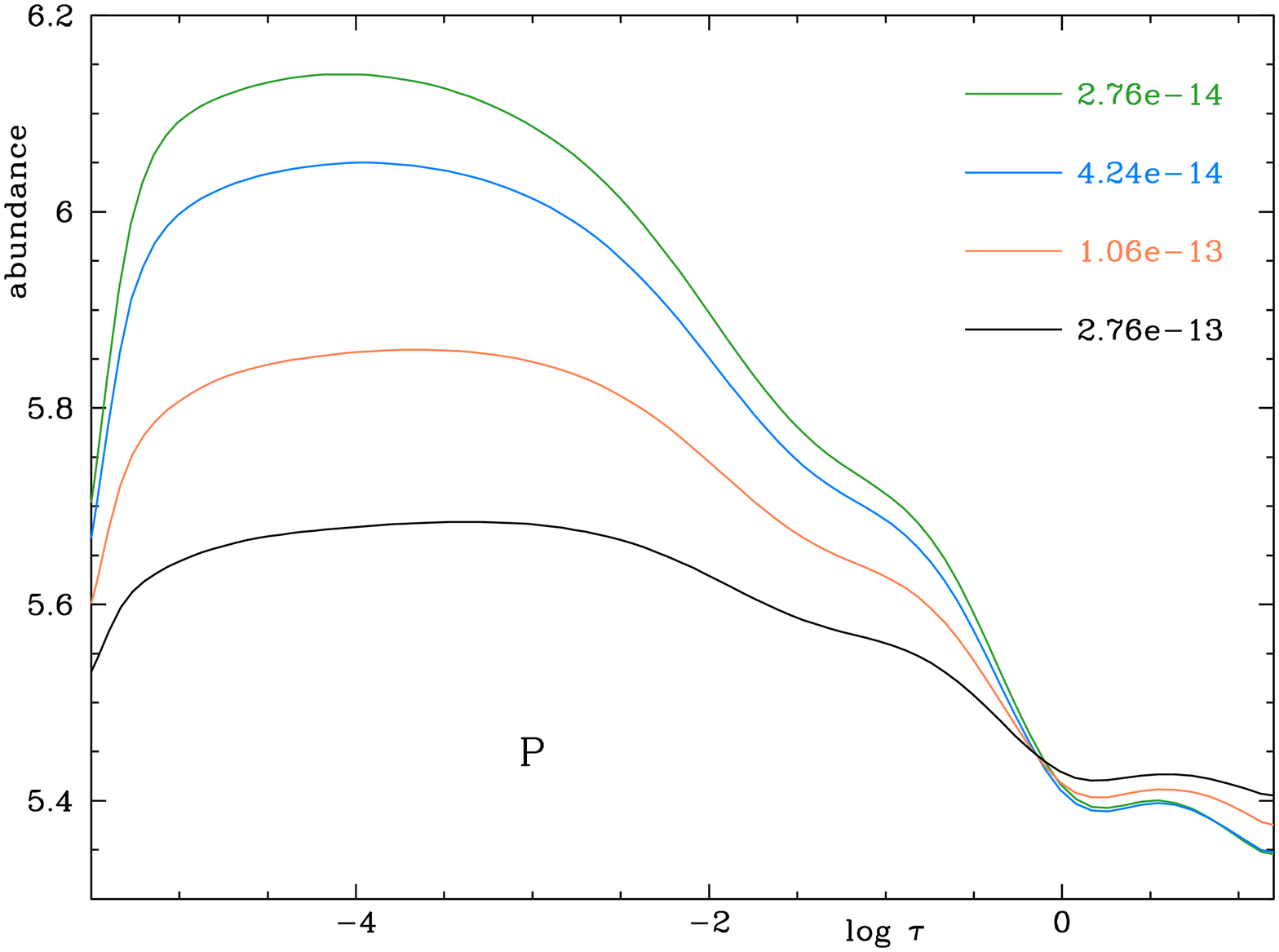}
\caption{
  {\bf a)} Stationary time-dependent chemical stratifications of Cr in magnetic
  fields of various absolute strengths (given in kG), inclined by $75\degr$
  with respect to the vertical. The stellar atmosphere is characterized by
  $T_{\rm eff} = 12000$\,K, $\log g = 4.0$ and a mass loss rate of $8.5\times10^{-15}$
  solar mass per year. The curves show abundances normalized to $[H] = 12.00$
  vs. $\log \tau_{5000}$.
  {\bf b)} Same stellar atmosphere, but stationary time-dependent chemical
  stratifications of Fe in a 17.5\,kG magnetic field as a function of the
  angle with respect to the vertical.
  {\bf c)} Stationary time-dependent chemical stratifications of P in the
  case of zero magnetic field as a function of the mass loss rate. The
  results pertain to a stellar atmosphere with $T_{\rm eff} = 12750$\,K and
  $\log g = 3.7$.}
\label{fig_1}
\end{centering}
\end{figure*}

\section{Time-dependent atomic diffusion in the presence of magnetic fields}
\label{diffus}

Ever since the ground-braking papers of \citet{VauclairVaHaPe1979},
\citet{MichaudMiChMe1981}, \citet {AlecianAlVa1981} and of
\citet{BabelMichaud1991a, BabelMichaud1991b}, there can be no doubt that the
magnetic fields found in CP stars control the build-up of stratified abundances.
It follows that the variation of inclination and of strength of the field over
the stellar surface -- be it simply bipolar or multipolar -- will invariably
lead to abundance inhomogeneities, both vertical and horizontal. In a series of
papers, Alecian and Stift have modeled this process to an increasing degree of
sophistication, starting with the effect of Zeeman splitting on accelerations
\citet{AlecianAlSt2002, AlecianAlSt2004}, translating accelerations to diffusion
velocities \citep{AlecianAlSt2006i}, establishing equilibrium solutions with
zero particle flux \citep{AlecianAlSt2007g} and demonstrating what chemical
stratifications would look like in a star permeated by a dipolar field
\citep{AlecianAlSt2010}. The 3D-modeling of a star with a non-axisymmetric
field geometry shows warped pseudo-rings of enhanced and depleted abundances,
greatly varying with depth in the atmosphere \citep{AlecianStift2017}.

In another approach to diffusion in CP stars \citep{AlecianAlStDo2011}, mass
conservation is included in the calculations; iterations start from vertically
constant abundances and continue until a stationary solution has been reached.
Subsequent modeling with up to 7 chemical elements ranging from Ti to Ni showed
how sensitively abundances depend on field inclination, even in the presence of
quite moderate field strengths \citep{StiftStAlDo2013, StiftAlecian2016}. A
certain degree of self-consistency is achieved by updating the atmospheric model
after a few iterations, a procedure that has also been applied to the equilibrium
solutions. Finally, the last years have seen the introduction of mass loss which,
not unexpectedly, has been shown to be able to greatly modify chemical stratification
profiles \citep{AlecianStift2019}.

Being a slow process, atomic diffusion is very sensitive to mixing motions.
The neglect of these in diffusion calculations is justified for ApBp stars
where observations indicative of abundance stratifications confirm the assumed
stability. On the other hand one cannot exclude that weakly turbulent layers
may at times exist in these atmospheres and some physical processes are also
missing in diffusion calculations such as for example NLTE effects -- see Sec.\,3
of \citet{Alecian2015} for a non-exhaustive list of missing processes. Still,
a number of general conclusions can be drawn from almost 4 decades of dedicated
efforts at ApBp star atmospheres:
\begin{enumerate}
\item Atomic diffusion leads to vertical abundance inhomogeneities.
\item Horizontal magnetic fields of virtually any strength greatly influence the
      resulting stratification profile at various atmospheric levels -- see e.g.
      Cr stratifications in fields of $75\degr$ inclination (Figure\,\ref{fig_1}a).
      Even a 1\,kG field leads to an abundance increase by about 1\,dex in the
      outermost layers.
\item From Figure\,\ref{fig_1}a it is evident that weak fields play a role only in
      the upper layers. It needs strong fields to affect abundances deep in the
      atmosphere.
\item Field inclination has a greater effect on chemical stratifications in the
      outer layers than field strength
\item Abundance profiles cannot in general be approximated by the popular step
      function The Fe stratifications in a 17.5\,kG field (Figure\,\ref{fig_1}b)
      for example reveal the existence of cloud-like structures.
\item In magnetic CP stars with moderate to strong fields one cannot therefore
      expect any kind of globally constant chemical stratifications.
\item Mass loss modifies the vertical abundance profiles, suppressing diffusion
      almost completely at high mass loss rates -- see the dependence on mass 
      loss of the stratifications of phosphorus in a zero field case
      (Figure\,\ref{fig_1}c).
\end{enumerate}
All the stratifications shown in these figures and later ones in this paper
have been obtained with the help of the {\sc CaratMotion} code described by
\citet{AlecianStift2019}. Based on the conclusions given above and on a wealth
of further numerical results, we may now proceed to analyze some aspects of the
(Zeeman) Doppler mapping approach to CP stars as practiced over the past 20 years.

\section{Abundance mapping: assumptions and simple tests}

There is no need to carry out another comparison of the respective virtues of the
Tikhonov versus the maximum entropy functionals in the least-squares minimization
problem at the basis of ZDM. Ever
since the paper by \citet{PiskunovPiKo2002}, (Zeeman) Doppler mapping of CP stars
has been monopolized by the {\sc invers} family of closely related codes, so that
we may justifiably restrict our discussion to Tikhonov regularization and these
codes. Thankfully, the numerous simplifying assumptions made in order to facilitate
(and indeed to make possible) ZDM of magnetic CP stars have been laid out in detail
by KP02. Chemical spots are characterized by vertically constant abundances, and
{\sc invers} sticks to this clearly unphysical approximation for all CP stars, even
when in HD\,3980 silicon seemingly becomes locally as abundant as hydrogen throughout
an atmosphere whose overall temperature and pressure structure however is given by
(much lower) abundances averaged over the star \citep{NesvacilNeLuShetal2012}.
\citet{StiftStLeCo2012} have shown that the use for spectrum synthesis of such mean
atmospheres for every location on spotted stars, regardless of the actual local
chemical abundances, can lead to serious errors in the recovered abundance maps.

\begin{figure*}
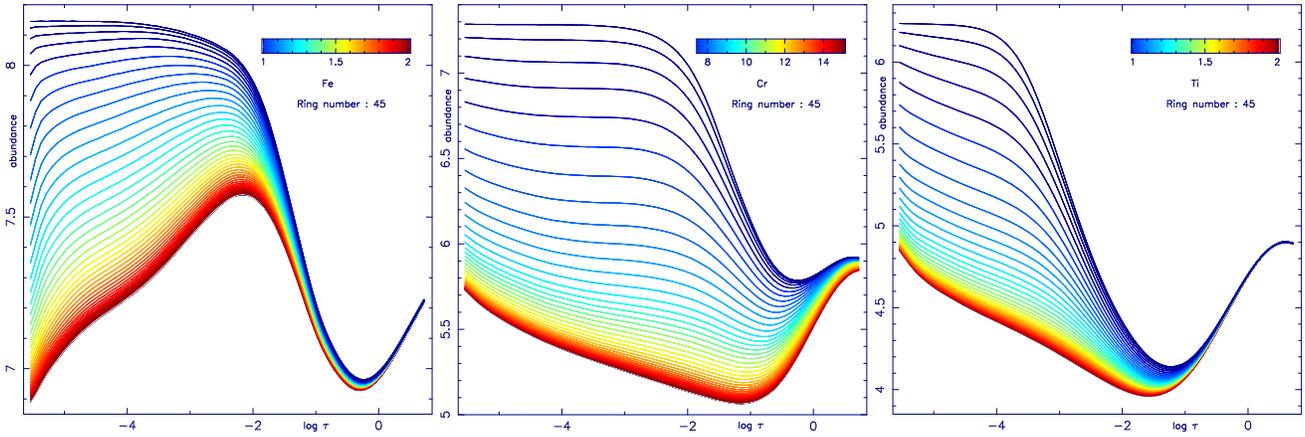

\begin{centering}
\includegraphics[width=57mm,height=57mm, angle=270]{lati_str.1012700577.Fe.45.eps}
\includegraphics[width=57mm,height=57mm, angle=270]{lati_str.1012700591.Cr.45.eps}
\includegraphics[width=57mm,height=57mm, angle=270]{lati_str.1012700577.Ti.45.eps}
\caption{
  Same atmosphere as in Figure\,\ref{fig_1}a,b.
  {\bf a)} Chemical stratifications of Fe in the atmosphere of a star with
  a centered dipole of 2\,kG polar field strength. Abundances along a meridian
  are color-coded according to the local field modulus and plotted versus the
  logarithm of optical depth. Same stellar atmosphere as in Figures\,\ref{fig_1}a,b.
  {\bf b)} Same as before but for Cr and a polar field strength of 15\,kG.
  {\bf c)} Same as before but for Ti and a polar field strength of 2\,kG.}
\label{fig_2}
\end{centering}
\end{figure*}

Fundamental physical questions are raised by results published for HR\,3831
\citep{KochukhovKoDrRe2003} with oxygen more abundant than hydrogen in an
exceedingly small spot on and Ca locally almost as abundant as helium --
remember that their oxygen and calcium spots stretch throughout the
atmosphere unstratified. The problem
of horizontal pressure equilibrium arising from the variations in the local
atmospheric structure has never been addressed in ZDM studies, although one
can immediately glean from published magnetic maps that in many places there
is no strong vertical magnetic field that could possibly stabilize the local
atmospheres \citep{StiftLeone2017a}. Why have stratifications as established
empirically for many CP stars since the 1990s -- albeit assumed globally
constant and mainly corresponding to step functions, see the comprehensive
presentation by \citet{Ryabchikova2014} -- in no way entered the inversion
procedure? No real simulations have ever been published as to how locally
variable stratifications will influence ZDM results based on vertically
constant local abundances. \citet{KochRyab2018} who constitute the sole
exception only look at a simple centered dipole.

Apart from the restriction that abundances in spots be unstratified, the
numerical tests devised for the mapping of abundance spots display a strange
lack of sophistication as to the shapes of the horizontal abundance
inhomogeneities assumed in \citet{KochukhovKoPi2002} and in \citet{Kochukhov2017}.
Just look at the Si, Cr and Fe maps of $\alpha^2$CVn \citep{KochukhovKoPiIlTu2002},
published almost simultaneously in the same series of papers. A star featuring
2 or 3 high-contrast spots of vertically constant abundance, well distributed
in both longitude and latitude, is an entirely artificial construct and there
is no guarantee nor is it even likely that it would correspond to what one
has to expect, even if exact theory were involved. Neither does this constellation
of spots bear the remotest resemblance to the numerous abundance maps of
$\alpha^2$CVn published over the years, even less so to the Ca and Fe maps of
HR3831 \citep{KochukhovKoDrPiRe2004}. In KP02's original tests, only 2 abundance
values were involved, in this respect very similar to the black and white values
in the famous ``Vogtstar'' example. The latter however is much more complex,
consisting of 4 letters with fine structure that Kochukhov's spots are completely
lacking. Still, despite adopting inclination, rotation, magnetic geometry, etc.
that are optimum for Doppler mapping, the results are somewhat disappointing. A
close look (better in color, not B/W) reveals that the quoted average error of
0.04\,dex is meaningless, reflecting a very specious statistical Ansatz; spurious
structures attain amplitudes of up to $\pm 0.5$\,dex. To summarize, the tests of
ZDM presented so far are based on artificial scenarios where diffusion is absent,
spots are of high contrast, small in number, largish, perfectly symmetric and well
distributed in longitude and latitude. In other words, they are entirely alien to
the complex world of magnetic CP stars.

\section{1D to 3D regularization in magnetic CP stars:
  abundances and magnetic fields}
\label{1Dstrat}

The indiscriminate use of physics-free regularization in the analysis of CP
stars can lead to strange results, even outside the strict realm of ZDM. In
their attempt to determine global 1D stratifications of chemical elements in
the atmosphere of HD\,133792, \citet{KoTsRyMaBa2006} claimed to have devised a
technique that ``for the first time allowed us to recover chemical profiles
without making a priori assumptions about the shape of chemical distributions.''
The Tikhonov approach chosen for the analysis of HD\,133792 minimizes the sum
of the squares of the vertical abundance gradients with optical depth, a purely
mathematical constraint which can easily be shown to have nothing to do with
theoretically modeled chemical stratifications in a magnetic CP star. Look for
example at the stratifications predicted for simple centered dipoles of 2\,kG
and 15\,kG polar strength respectively. Based on a grid of chemical profiles
for field strengths from 1\,kG to 20\,kG and field directions from vertical to
horizontal, one can establish the vertical abundance structure along a meridian
(for a dipole aligned with the rotational axis) or the equator (for a dipole
lying in the equatorial plane) as plotted in Figures\,\ref{fig_2}a,b,c. Even in
the case of a low contrast, weak field -- the field modulus ranges from 1 to
2\.kG -- Fe abundances in the upper layer diverge by more than 1\,dex. As
expected, we see a much greater effect reaching 2\,dex for Cr in fields ranging
from 7.5 to 15\,kG. If one can discern anything bearing resemblance to 
a ``transition region'' between low and high abundances in
the Cr stratifications, its extension, shape and slope greatly varies all along
the meridian, at variance with the assumptions underlying the 3D regularization
functional used by \citet{Rusomarov2016} (see below). It becomes immediately
obvious that the adoption of a globally constant vertical abundance profile in
the analysis of HD\,133792 disregards the complex astrophysical
reality. From a comparison between our Figures\,\ref{fig_2}a,b,c and Figures\,3
and 5 of \citet{KoTsRyMaBa2006} it also transpires that in stark contrast to the
assertions made, completely {\em unphysical} a priori assumptions have silently
been introduced, concerning the shape of the chemical distributions, viz. a very
special kind of smooth step function which asymptotically reaches the solar
abundance both in the deepest and in the outermost layers. See
\citet{StiftStAl2009s} for further discussions.

\begin{figure*}
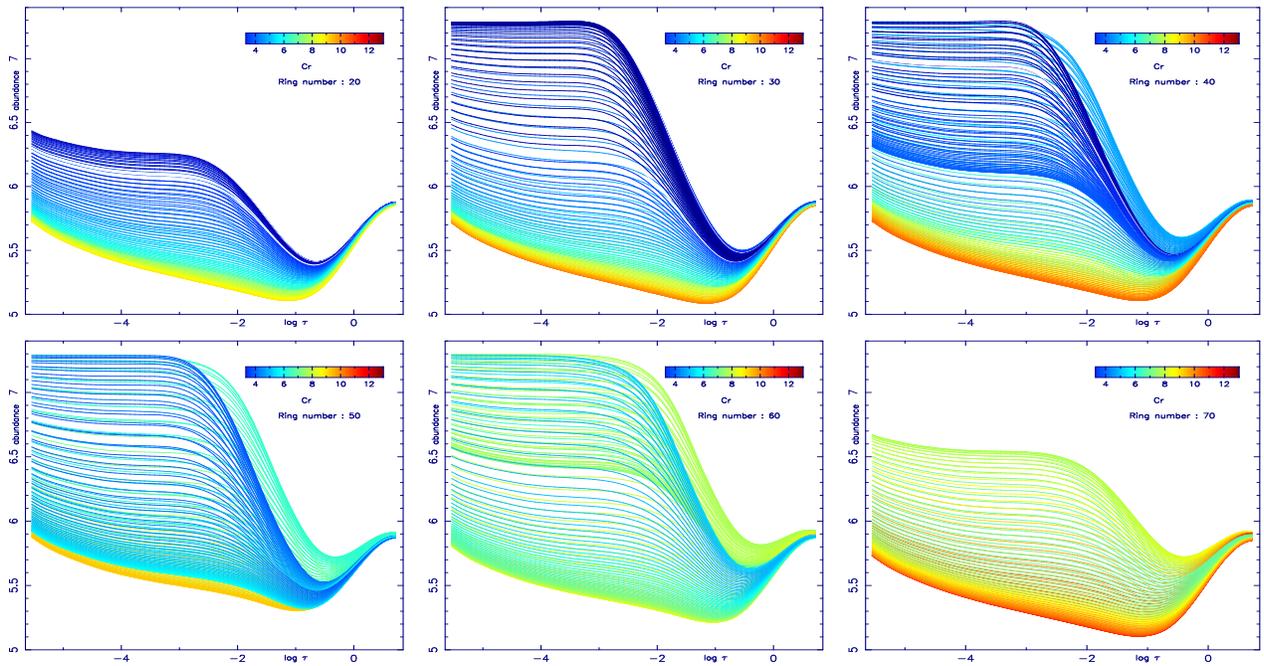

\begin{centering}
\includegraphics[width=44mm, height=55mm, angle=270]{lati_str.1012746213.Cr.20.eps}
\includegraphics[width=44mm, height=55mm, angle=270]{lati_str.1012746213.Cr.30.eps}
\includegraphics[width=44mm, height=55mm, angle=270]{lati_str.1012746213.Cr.40.eps}
\includegraphics[width=44mm, height=55mm, angle=270]{lati_str.1012746213.Cr.50.eps}
\includegraphics[width=44mm, height=55mm, angle=270]{lati_str.1012746213.Cr.60.eps}
\includegraphics[width=44mm, height=55mm, angle=270]{lati_str.1012746213.Cr.70.eps}
\caption{
  Chemical stratifications of Fe in the atmosphere of a star with a magnetic
  geometry similar to that of HD\,154708. The stratification profiles are
  color-coded according to the local field moduli which range between 3.35 and
  12.89\,kG; they are plotted versus the logarithm of optical depth. Ring
  numbers 20, 30, 40, 50, 60, and 70 correspond to latitudes $-50\degr$, $-30\degr$,
  $-10\degr$, $+10\degr$, $+30\degr$, and $+50\degr$. Same stellar atmosphere as
  in Figure\,\ref{fig_1}a,b}. 
\label{fig_3}
\end{centering}
\end{figure*}

Both in the mapping of 2D horizontal abundance maps and in the derivation of
vector magnetic field maps of CP stars, Tikhonov regularization is presently
the only constraint in use. Unlike the ``Vogtstar'' case which is conveniently
devoid of physical meaning, stratified abundances are governed by processes
involving magnetic fields (have a look again at Figures\,\ref{fig_2}a,b,c), but
this is not reflected in the functionals employed in the {\sc invers} family
of codes. \citet{KochukhovKoWa2010} give one example in their eq.\,(5), where
they minimize simultaneously the double sum of the squared differences between
all combinations of 2 magnetic vectors and all combinations of 2 abundances of
the chemical elements considered in the inversion. One would be hard pressed
to devise a more arbitrary and less physically motivated, yet mathematically
expedient regularization: neither does it take into account the properties of
truly multipolar field geometries -- not just dipole or dipole plus quadrupole
--  nor does it in the least reflect the dependence of chemical stratifications
on field direction and field strength. There are no tests to be found in the
literature that would demonstrate how this regularization could possibly ensure
the correct recovery of a magnetic field with poloidal and toroidal components
of spherical harmonics degree and order of 10 or more. Let us recall that the
tests presented by KP02 deal exclusively with a dipole of 8\,kG\,(!) polar
strength and a few models with an added axisymmetric quadrupole, of 8\,kG
strength again. Tests based on these huge field strengths and devised in a way
that ``magnetic geometry, rotational velocity and inclination collaborate to
maximize the magnetic variability of the spectral line profiles and represent an
ideal combination for the application of MDI'' cannot provide answers to this
question.

Without further discussion, \citet{KochukhovKoLuNeAl2014} have introduced a
new penalty function in the context of inversions where spherical harmonics
are fitted to the observed magnetic field. They chose the expression
$\Sigma_{l,m}~l^2\,(\alpha^2 + \beta^2 + \gamma^2)$ where $\alpha, \beta$ and
$\gamma$ characterize contributions of the radial poloidal, horizontal
poloidal, and horizontal toroidal magnetic field components, respectively.
Again, there are neither tests nor theoretical considerations to show that
this particular penalty function would ensure correct magnetic maps. A mere
2 years later, \citet{RusomarovRuKoRyIl2016} presented yet another penalty
function $\Sigma_{l,m}~l^2\,(\alpha + \beta + \gamma)^2$; this surprising change
was not accompanied by a single argument nor was it buttressed by tests similar
to those published by KP02. This leaves the world with at least 3 penalty
functions, all of which constitute largely or even entirely untested ad hoc
constructs. We will demonstrate later in this paper that these penalty
functions fail to lead to physically feasible field geometries in strongly
magnetic CP stars. This failure likely extends to weak-field CP stars.

The application of Tikhonov regularization to 2D abundance maps neglects the
undisputed fact that stratifications are related to the local magnetic field.
Figure\,\ref{fig_3} reveals the diversity of chemical profiles in a star with a
magnetic field geometry similar to that of HD\,154708 \citep{StiftStHuLeetal2013}.
ZDM in its present form tries to fit integrated line profiles (originating from
vertically and horizontally varying stratifications) to integrated line
profiles assumed to emanate from horizontally varying but vertically constant
abundance profiles. At the same time the unstratified profiles forced onto the
vertically variable profiles have to conform to the artificial restraint of
minimum horizontal abundance gradients. Only simulations based on theoretical
stratifications could give us hints as to the interpretation of the results
of such a procedure.

\subsection{3D abundances and an improper regularization}
\label{3Dstrat}

Kochukhov et al. (2009, JD4 @ IAU GA XXVII) and later \citet{Rusomarov2016}
in section\,2.3 of his PhD thesis -- submitted as \citet {RusomarovRuKoRy2016}
to A\&A but never accepted for publication in any
journal\footnote{http://urn.kb.se/resolve?urn=urn:nbn:se:uu:diva-278534}
-- attempted to derive 3D abundance profiles of Fe in the atmospheres of
$\Theta$\,Aurigae and of HD\,24712 respectively. In spite of what had already
been known for some time about stratifications in Ap star atmospheres (see
e.g. \citealt{AlecianAlSt2010}), Rusomarov adopted the regularization scheme
previously devised by Kochukhov that minimizes all possible differences
between the totality of idealized local stratification profiles; for numerical
convenience the latter are assumed to be step-like. In the Kochukhov-Rusomarov
picture, the depth-dependence of the abundance profiles is a function of 4
parameters: abundance in the upper atmosphere $\epsilon_{\rm up}$, abundance
deep in the atmosphere $\epsilon_{\rm low}$, position $d$ and width $\delta$ of
a transition region lying somewhere in between. From these 4 parameters,
5 quantities are formed whose square sums are minimized simultaneously, viz.
$\Sigma\,|\Delta\epsilon_{\rm up}|^2$,
$\Sigma\,|\Delta\epsilon_{\rm low}|^2$,
$\Sigma\,|\Delta\,d|^2$,
$\Sigma\,|\Delta\delta|^2$,
$\Sigma\,(\epsilon_{\rm up} - \epsilon_{\rm low})^2$.
No theoretical basis is offered for this particular regularization, neither in
section\,2.3 of the thesis nor in the A\&A submission. One also acutely misses
tests that would show how and under which conditions the proposed ZDM approach
could be able to recover complex 3D chemical profiles as for example predicted
by diffusion theory. Instead of finding a carefully devised, executed, described
and discussed set of tests we are faced with a simple (unspoken) extension to
the old KP02 belief ``that the code can be successfully applied to the imaging
of global stellar magnetic fields and abundance distributions of an arbitrary
complexity''. Apparently the exact form of regularization functionals used in
ZDM does not matter and therefore does not have to be argued; whenever
regularization is presented to the astronomical community as Tikhonov-like, it
seemingly is expected to be accepted without further questioning.

The approach chosen by Kochukhov and by Rusomarov is clearly at variance with
the diffusion models of \citet{AlecianAlSt2007g} which show that chemical
profiles strongly depend on magnetic field direction and strength. This holds
true not only for equilibrium solutions but also for the more realistic
time-dependent results, even though the final stationary solutions may deviate
from the solar abundances adopted at the start of the calculations in a sense
opposite to what is found for equilibrium solutions \citep{StiftAlecian2016}.
Figures\,\ref{fig_2}a,b,c have already revealed that one is faced with
considerable variations of $\epsilon_{\rm up}$ but also in the position $d$
and width $\delta$ of the transition region in a field geometry as simple
as a centered dipole, even at fairly modest field strength. Much enhanced
variations are found in another geometry of rather moderate complexity,
illustrated by Figure\,\ref{fig_3}. With local field moduli ranging between
3.35 and 12.89\,kG, this allegedly ``unspecified eccentric dipole with
an unusually high contrast'' (in fact representing the magnetic geometry of
HD\,154708) -- a mere factor of 3.8 as compared to 5.6 derived
for HD\,32633 by \citet{SilvesterSiKoWa2015} and a staggering 11.5 for
HD\,119419 according to \citet{RusomarovRuKoLu2018} -- helps to convey a
reasonably conservative estimate of how non-homogeneous, strong magnetic
fields can lead to an impressive variety of chemical stratifications over
the stellar surface.

From these plots and from stratifications predicted for many other
geometries it follows that forcing artificial uniformity onto idealized
local step-functions and then minimizing abundance differences at all
costs cannot possibly provide valid insights into the physics of CP stars.
The very peculiar, arbitrary and physics-free regularization functional
first chosen by Kochukhov and later taken up again by Rusomarov cannot
but lead to spurious solutions for the abundances, disconnected from the
true magnetic geometry. Such abundance maps rather imply an unnatural
magnetic geometry featuring small differences in field strength and even
smaller differences in field angle; they are entirely free from any
physical meaning,

\section{Magnetic fields of CP stars, through Maxwell to Alfv{\'e}n}
\label{field}

In the first ZDM mapping of a CP star based on all 4 Stokes parameters,
\citet{KochukhovKoBaWaetal2004} derived a vector magnetic field map of
53\,Cam, revealing absolute field strengths over the stellar surface ranging
from 4 to 26\,kG. Checking the consistency of their map with Maxwell's
equations, they found a large (44\%) magnetic flux imbalance. It is hard
to understand why 53\,Cam has never been reanalyzed in subsequent years --
in contrast to other stars first studied in the earliest days of ZDM such
as $\alpha^2$\,CVn or HD\,24712 -- despite these shortcomings and despite
the fact that present-day spectrographs would provide observational
material of much higher resolution and signal-to-noise ratio. At present
the true surface structure of 53\,Cam must thus be considered unknown.

In a later study, \citet{KochukhovKoWa2010} failed to provide details
as to a possible magnetic flux imbalance for $\alpha^2$\,CVn. The
approximation of their vector magnetic map with spherical harmonics
appears to be more satisfactory than for 53\,Cam, but the size of the
deviation from zero divergence of the discrete map remains unspecified.
A better situation seems to prevail for HD\,32633
\citep{SilvesterSiKoWa2015}, a star with a mean field modulus of about
8\,kG. In the magnetic field inversion based on all 4 Stokes parameters,
the {\sc invers} code was allowed to fit harmonics up to $l = 10$,
including poloidal and toroidal components. The radial and the horizontal
poloidal field components were determined separately, i.e. the
coefficients $\alpha_{\rm lm}$, $\beta_{\rm lm}$, and $\gamma_{\rm lm}$ of the
spherical harmonics expansion specifying the field geometry were allowed
to vary independently as in \citet{KochukhovKoLuNeAl2014}.

\begin{figure*}
\begin{centering}
\includegraphics[width=51mm, height=88mm, angle=270]{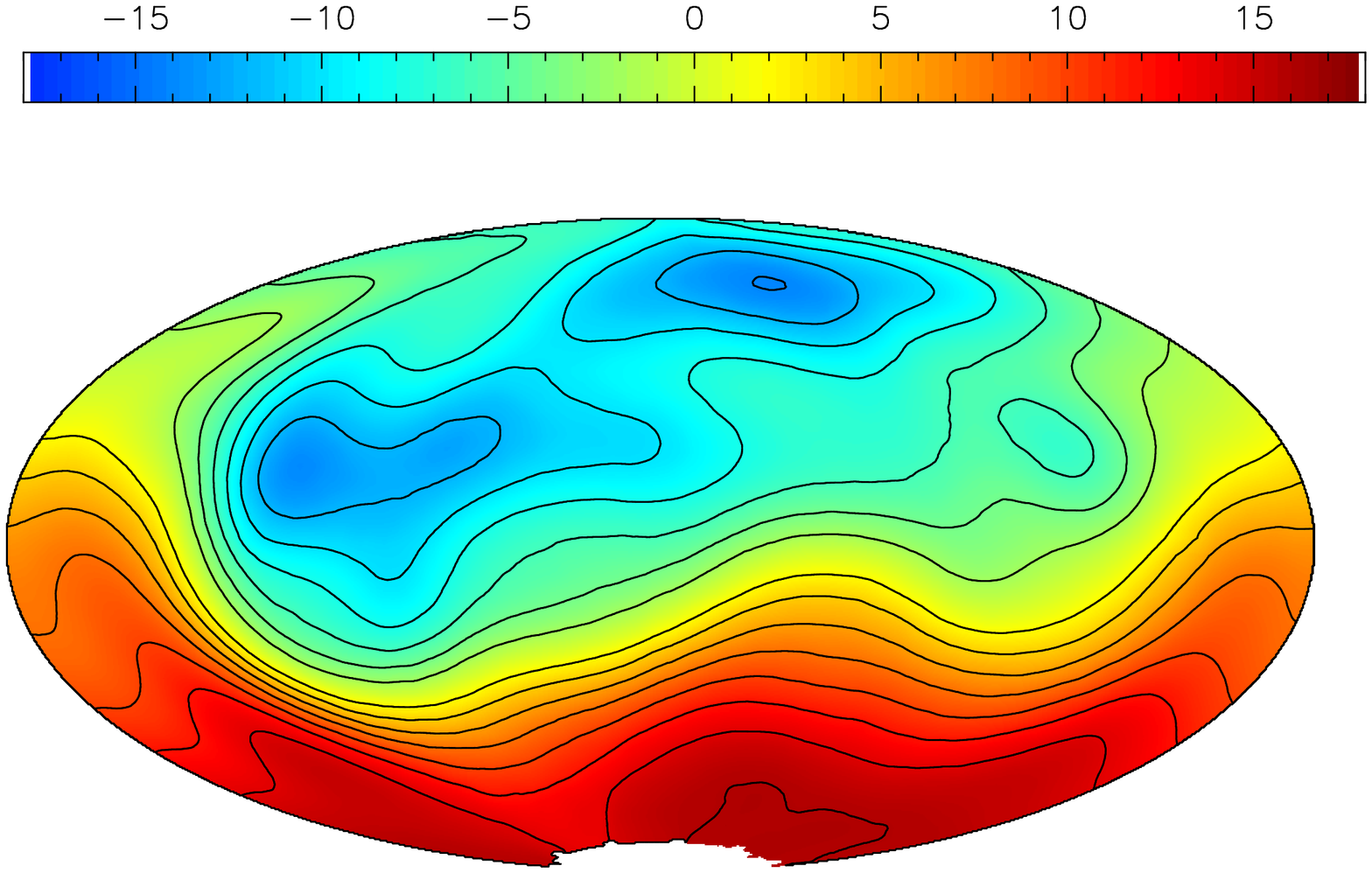}
\includegraphics[width=51mm, height=88mm, angle=270]{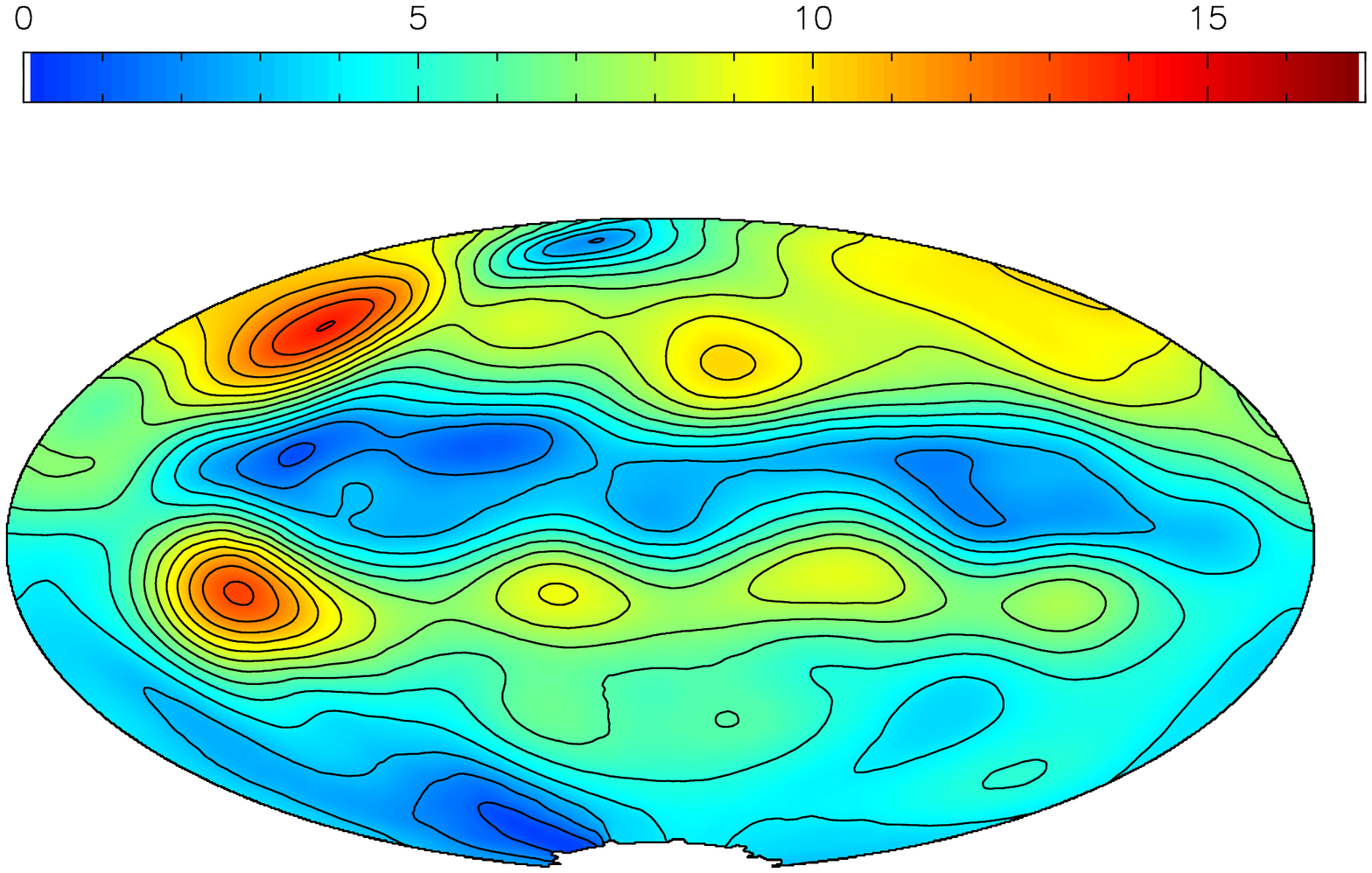}
\includegraphics[width=51mm, height=88mm, angle=270]{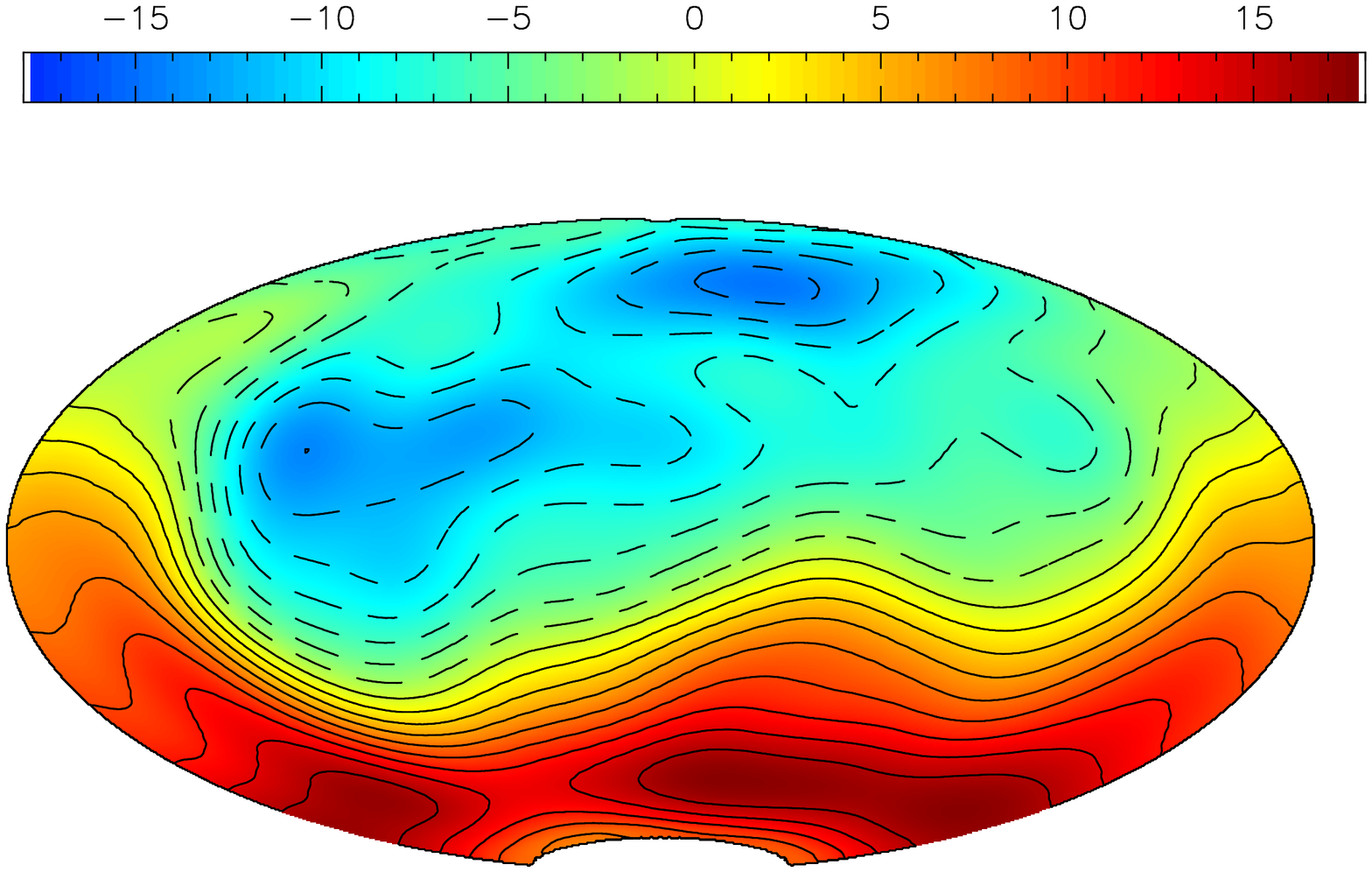}
\includegraphics[width=51mm, height=88mm, angle=270]{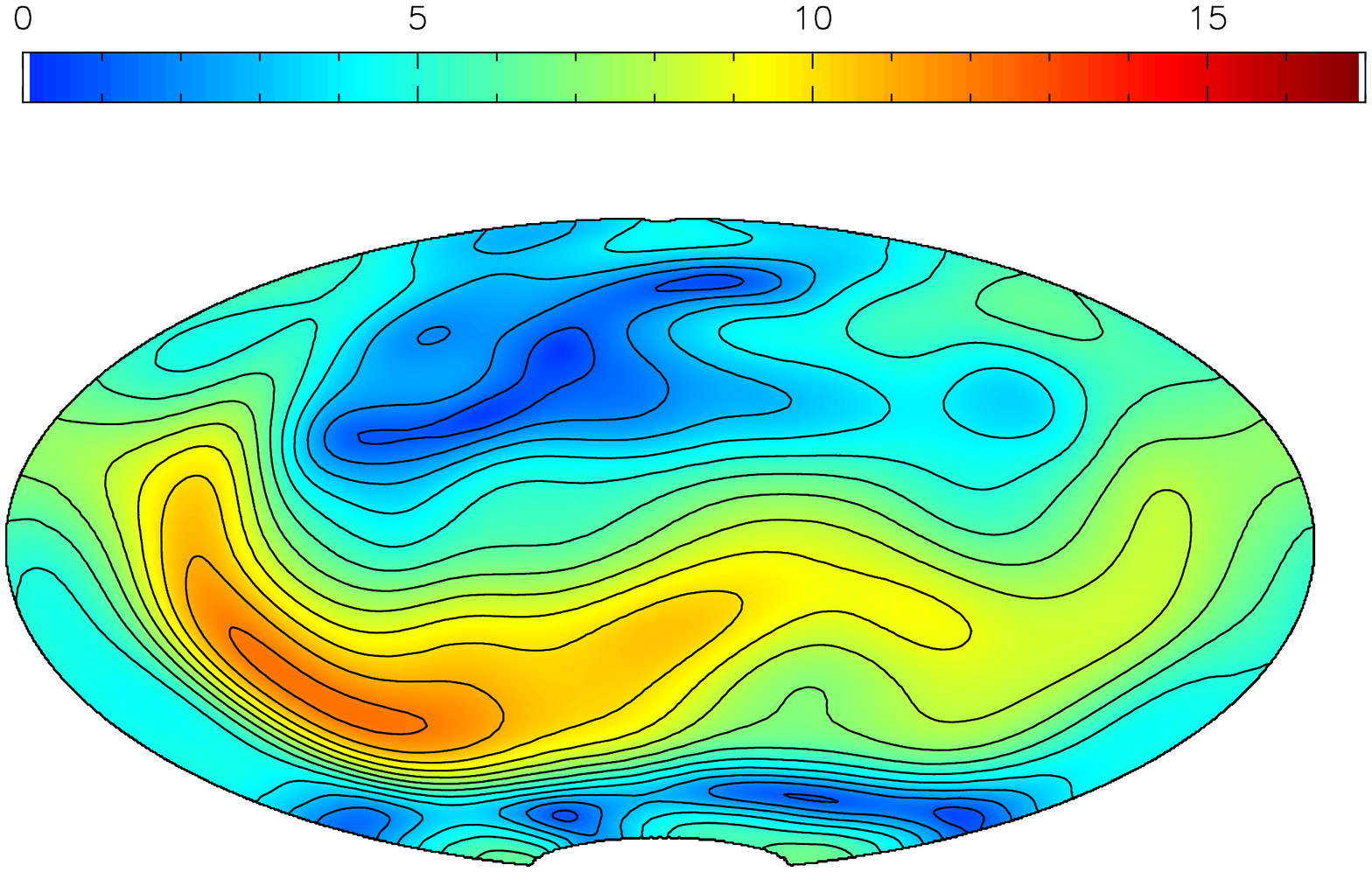}
\includegraphics[width=70mm, height=38mm, angle=0, origin=l]{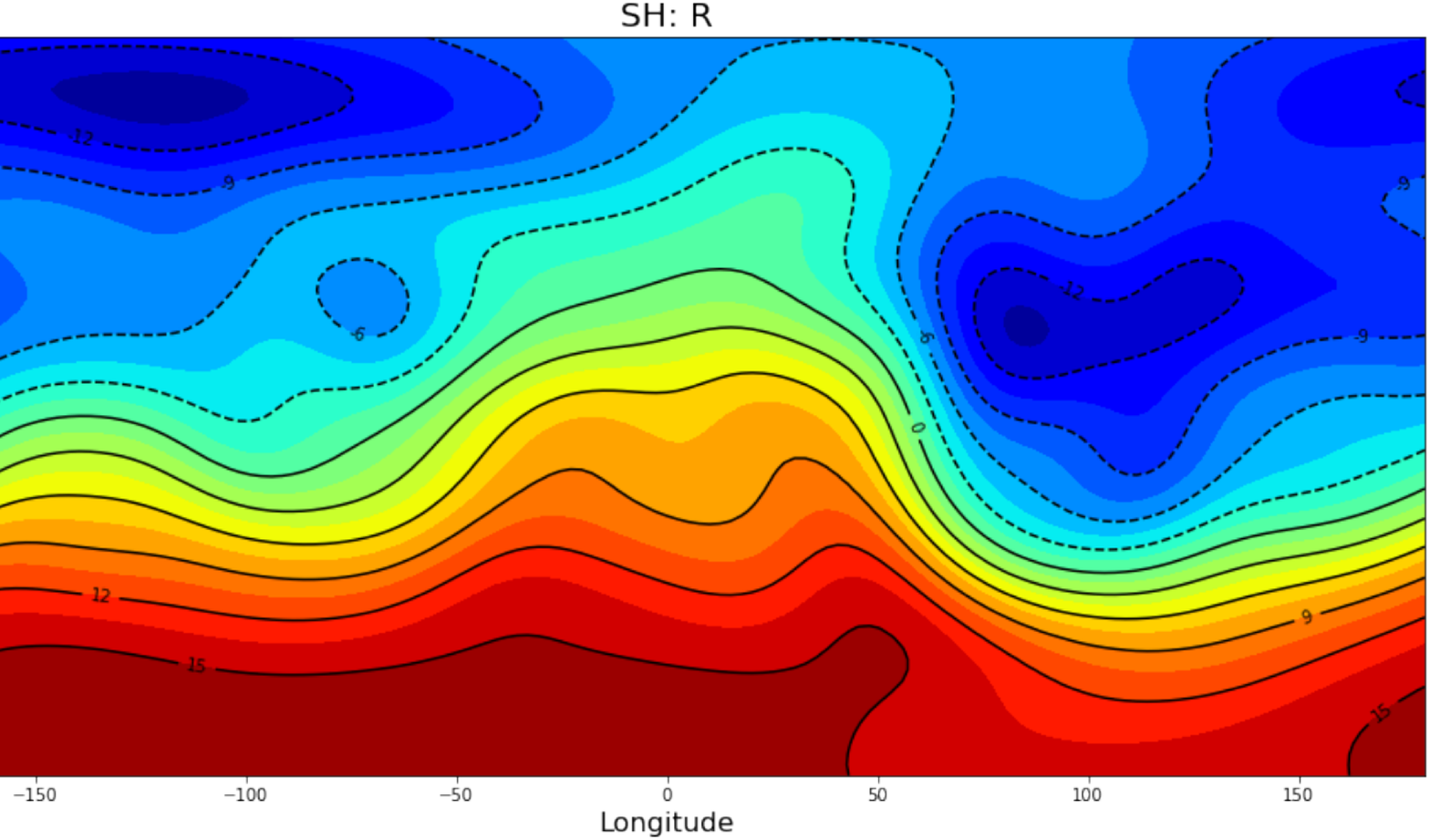}
\includegraphics[width=70mm, height=38mm, angle=0, origin=r]{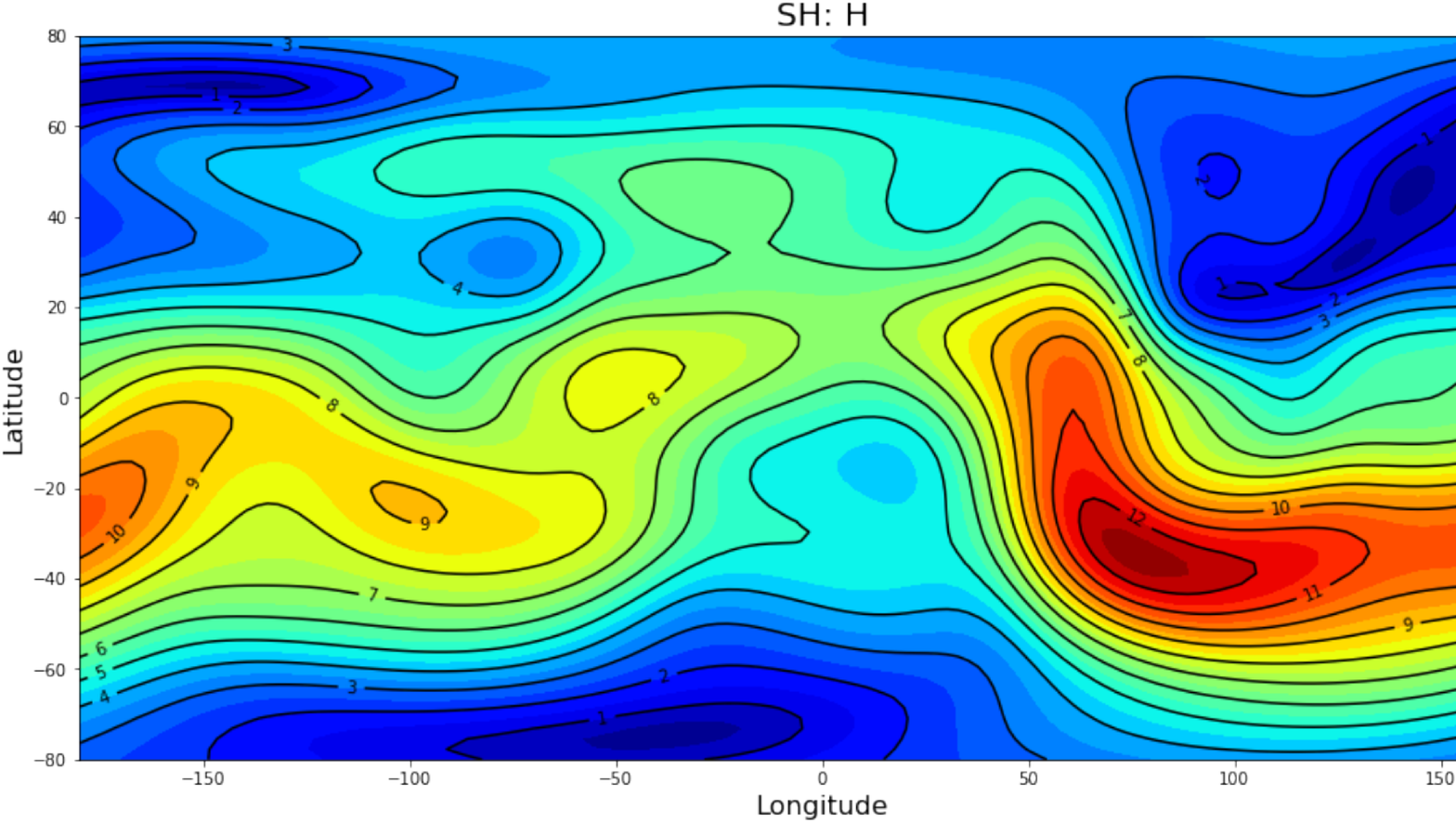}
\caption{
 {\bf a}) Radial magnetic field of HD\,32633 (left). From top to bottom :
  original field recovered from published plots, original field fitted
  with spherical harmonics, original field fitted with an independent
  code by C.D.~Beggan (British Geological Survey, Edinburgh). The phases
  differ by $180\degr$ between the two codes.
  {\bf b}) Horizontal magnetic field of HD\,32633 (right). From top to
  bottom: original field recovered from published plots, horizontal
  field derived from the radial field by assuming a potential poloidal
  force-free configuration, horizontal field derived similarly with the
  independent code by C.D.~Beggan.
  }
\label{fig_4}
\end{centering}
\end{figure*}

Enter Hannes Alfv{\'e}n and magnetohydrodynamics. It is well known
that in the solar corona magnetic structures have to be force-free,
$(\nabla \wedge B) \wedge B  = 0$, given the dominance of Maxwell stress
and/or magnetic pressure over gas pressure. Force-free fields are also
prominent in the solar chromosphere and they govern the structure of
sunspots (see e.g. \citealt{Tiwari2012}). The vertical magnetic fields
of sunspots only very rarely attain values of 4\,kG, so shouldn't the
field of HD\,32633 with regions of up to 17\,kG field strength as
derived by \citet{SilvesterSiKoWa2015} also qualify as force-free?
Spruit\footnote{http://www.mpa-garching.mpg.de/\~henk/mhd12.zip} (2017)
answers this in the affirmative and also points out that the construction
of a force-free field is not possible in terms of a boundary-value problem;
force-free fields must be understood in the context of the entire history
of the fluid displacements at their boundary. It follows that within the
framework of ZDM it is not possible to determine the shape of a force-free
configuration; the inversion of strong magnetic fields in CP stars must
needs be based on purely potential fields, ensuring $\nabla \wedge B = 0$
in addition to zero divergence.

\begin{figure*}
\begin{centering}
\includegraphics[width=42mm, height=88mm, angle=270]{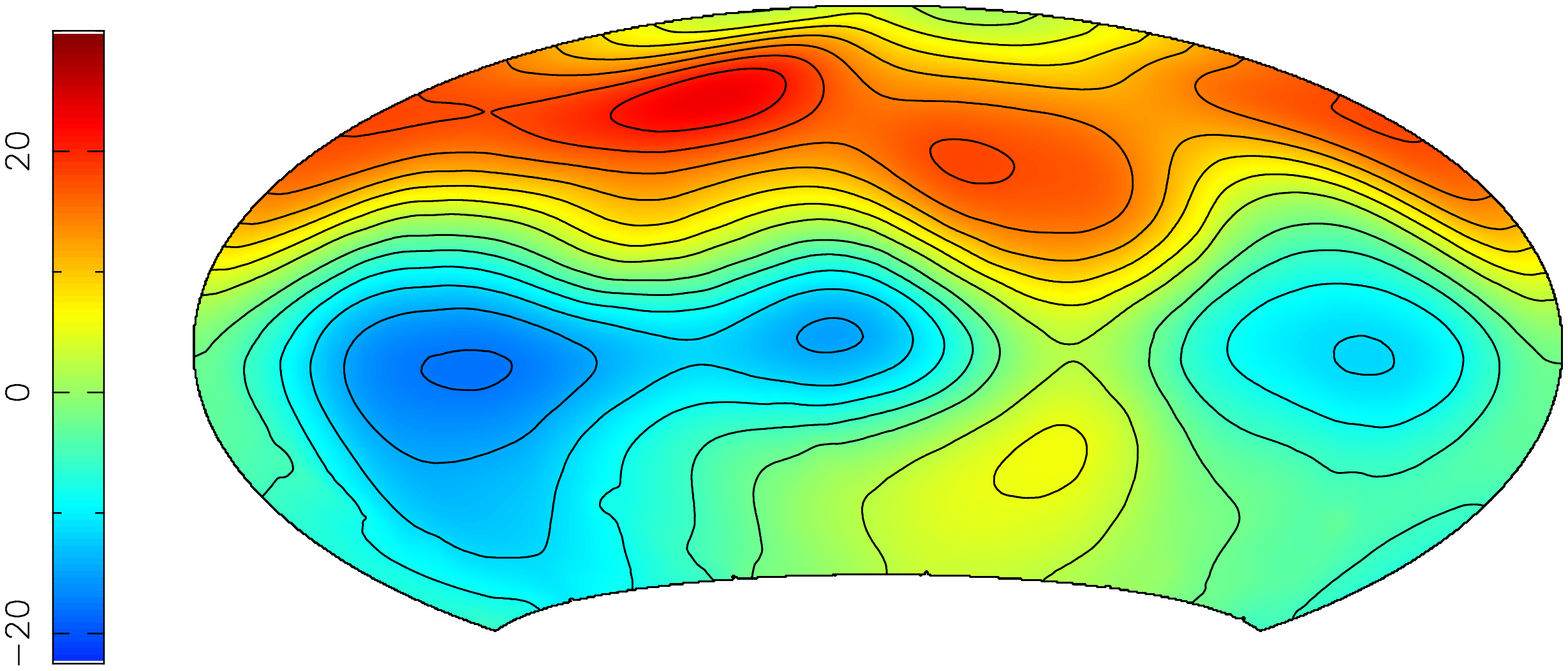}
\includegraphics[width=42mm, height=88mm, angle=270]{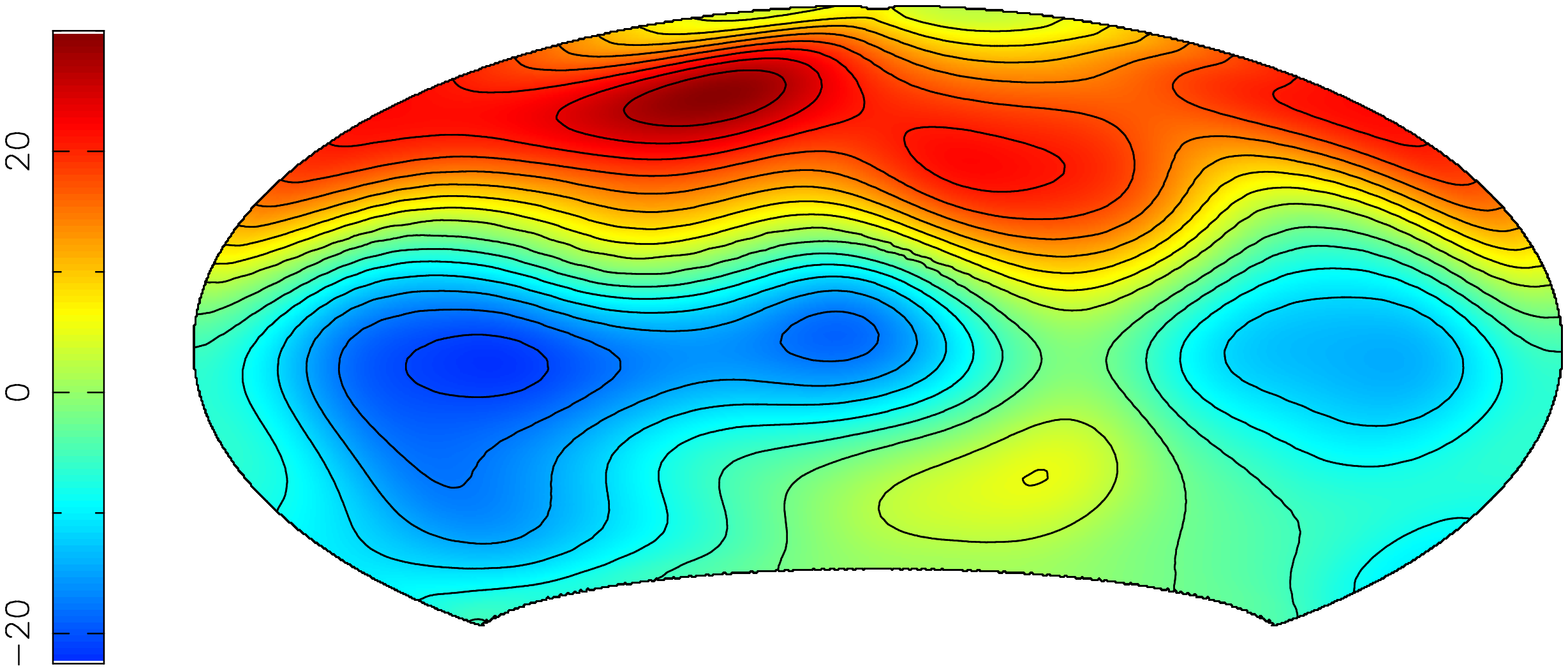}
\includegraphics[width=42mm, height=88mm, angle=270]{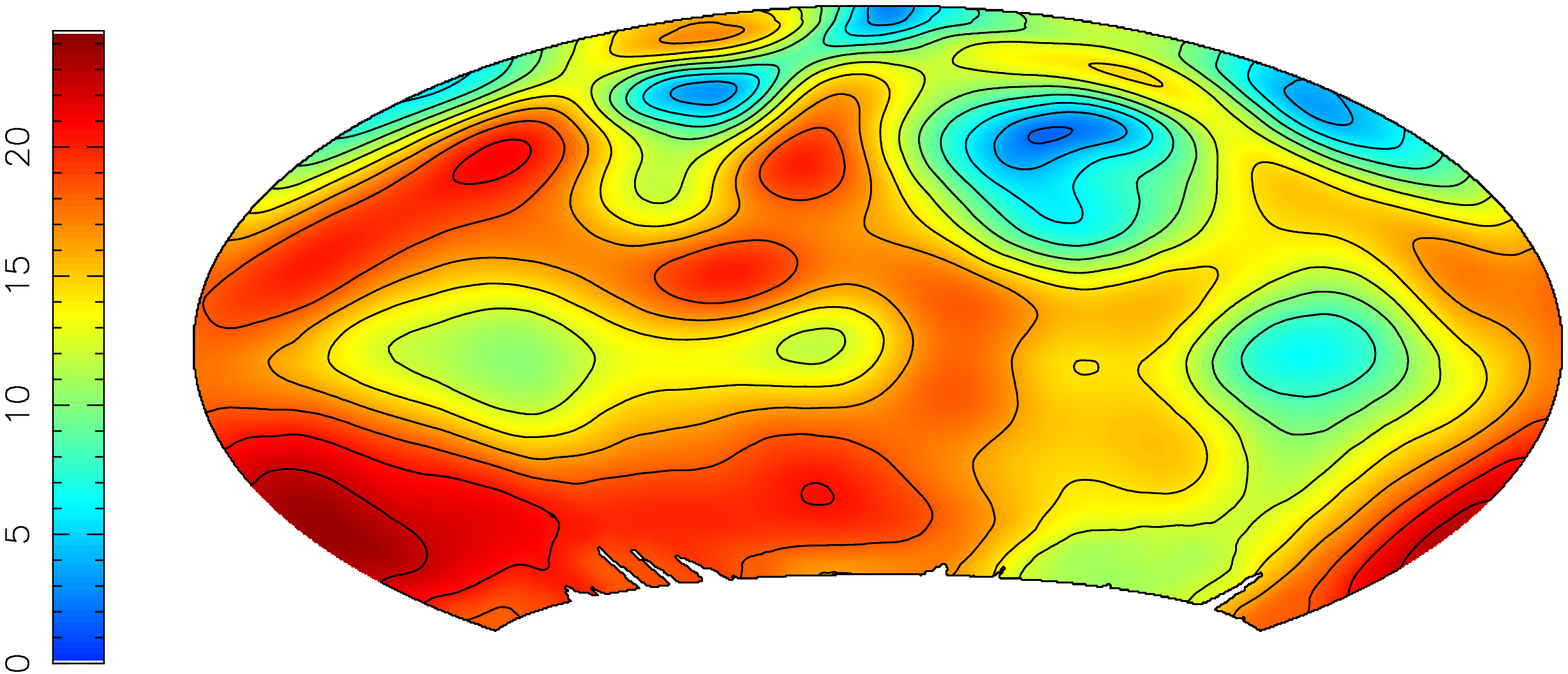}
\includegraphics[width=42mm, height=88mm, angle=270]{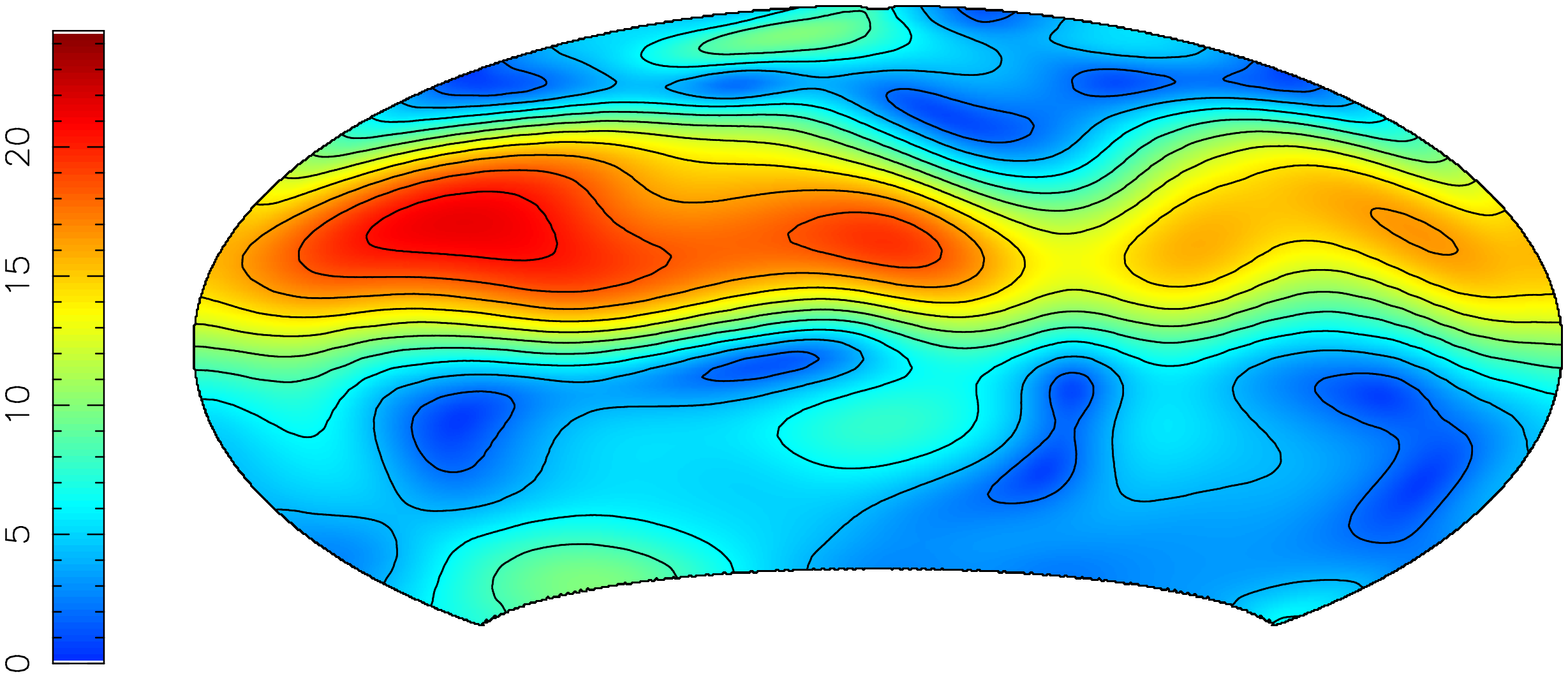}
\caption{
  {\bf a}) Original radial and horizontal magnetic field of HD\,119419 (left, top to bottom).
  {\bf b}) Corresponding force-free radial and horizontal magnetic field (right, top to
  bottom).
  }
\label{fig_5}
\end{centering}
\end{figure*}

Writing down $\nabla \wedge B = 0$ in spherical coordinates leads to
eqs. (12)-(14) of \citet{KochukhovKoBaWaetal2004} with all 3 components
of the current $J$ set to zero. These equations define the relation
between the components of the magnetic field vector. Since $B_{r}$ depends
on $B_{\phi}$ and on $B_{\theta}$,  it is not legitimate to determine
$\alpha_{\rm lm}$ independently from $\beta_{\rm lm}$ and $\gamma_{\rm lm}$.
Eqs. (1)-(3) of \citet{KochukhovKoLuNeAl2014} can possibly be applied
to very weak fields, but certainly not to stars like
HD\,32633  \citep{SilvesterSiKoWa2015},
53\,Cam    \citep{KochukhovKoBaWaetal2004},
HD\,75049  \citep{Kochukhovetal2015},
HD\,119419 \citep{RusomarovRuKoLu2018},
HD\,125428 \citep{RusomarovRuKoRyIl2016},
HD\,133880 \citep{KochukhovKoSiBaLaWa2017}
and probably also not to
49\,Cam             \citep{SilvesterSiKoRuWa2017}
and $\alpha^2$\,CVn \citep{SilvesterSiKoWa2014}.
In view of the field strengths observed, it perspires that {\em all} Zeeman
Doppler maps of CP stars with strong fields have been obtained with an incorrect
set of formulae. The published magnetic maps for the above-mentioned CP
stars are therefore {\em all} entirely spurious, and so are {\em all} the
abundances, since Zeeman splitting, Zeeman intensification and polarization
of the spectral lines are based on erroneous magnetic field values.
The situation is not quite so clear-cut with
stellar fields of more moderate strength but as Spruit (private communication)
points out, these are probably also stable on very long time scales so that
Ohmic diffusion has given the field in the atmosphere sufficient time to
relax to its lowest energy state, viz. a potential field.

\subsection{Published vs. force-free maps}
\label{HD32633}

Geophysicists modeling the earth's magnetism do not restrict the dissemination
of their results to colorful plots in journals and on web-pages, rather they
provide extensive tables, Fortran codes with hundreds of lines of data, notebooks,
etc... This allows fellow scientists to take advantage of the existing wealth of
models for further investigations; it also constitutes a useful check on the
integrity of data and models. No openness of this kind is encountered in ZDM,
quite to the contrary not even the coefficients of the spherical harmonics
expansion of any of the many CP stars analyzed has ever been made available
apart from those for 36\,Lyn \citep{Oksala_etal_2018}. Over the years, there
has been no way to obtain ZDM data in view of an {\em independent} assessment of
the published maps and of further analyses. Among others, this inaccessibility
of the ZDM data precludes the prediction (by scientists working on atomic
diffusion) of observable Stokes profiles, based on theoretical field-dependent
chemical stratifications unless close supervision is accepted. Kochukhov
verbatim: {\it ``I provide access to my DI methods in the context of
collaborative projects, in which I am involved at all stages from problem
formulation to eventual publications. In most cases I compute DI maps myself
using the data files prepared by collaborators.''} Being in themselves
legitimate, these wide ranging restrictions have led to the scientifically
highly undesirable situation that not only the ZDM maps, but even the methods
leading to these maps, have largely remained unassailable over the years
-- for the few exceptions see e.g. \citet{Stift1996}, \citet{StiftStLeCo2012},
\citet{StiftLeone2017b, StiftLeone2017a}.

Fortunately, things have changed recently, although neither journal editors
nor referees have insisted at last on making ZDM data routinely available. A
few transformations applied to the published ZDM maps suffice to recover
magnetic field geometries to a gratifying degree of reliability.
Figure\,\ref{fig_4}a (top) shows the radial field of HD\,32633 as reconstructed
from the maps published by \citet{SilvesterSiKoWa2015}, Figure\,\ref{fig_4}b
(top) the horizontal field. Below to the left we show a spherical harmonics
fit with $l = 1-9$ applied to the recovered map of the radial magnetic
field. The excellent agreement with the original plot is quite surprising,
given the many steps involved in the reconstruction. For all practical purposes,
the data underlying the Hammer projections in these plots are near enough to
the results obtained with the {\sc invers10} code to be used straightforwardly
in further analyses. We know that in the case of potential poloidal (and thus
force-free) fields, the horizontal field components can be derived directly
from the radial field (see e.g. \citealt{WinchWiIvTuSt2005}). For the necessary
calculations we have developed a new code, testing it with the International
Geomagnetic Reference Field \citep{IGRF13}. We note that the resulting map
(Figure\,\ref{fig_4}b middle) of the horizontal field -- force-free as required
in view of the strong magnetic fields involved -- is totally at variance with
the original map which represents a non-potential field with magnetic forces
that are not in balance. In an independent analysis of the original magnetic
maps, C.D.~Beggan (British Geological Survey, Edinburgh) has confirmed that
from the raw published radial field maps, one cannot reproduce the horizontal
or modulus plots shown in the paper (bottom of Figures\,\ref{fig_4}a,b).

We also had a look at HD\,119419 \citep{RusomarovRuKoLu2018} with its
magnetic field modulus reaching about 26\,kG and its spectacular 4 ``spots''
of very low values of the horizontal field. Strangely, here the spherical
harmonics fit to the radial field is somewhat unsatisfactory, the residuals
being almost 3 times as large as for HD\,32633. Although we feel unable to
explain this discrepancy, Figure\,\ref{fig_5}a shows a field map sufficiently
near to the original one about the equator and in the northern hemisphere
to make possible the desired further analyses. Proceeding as for HD\,32633
with our well tested code, we show that the published horizontal field
certainly is not force-free (Figure\,\ref{fig_5}b).

Similar analyses should prove straightforward for HD\,125428
\citep{RusomarovRuKoRyIl2016} and HD\,133880 \citep{KochukhovKoSiBaLaWa2017}.
In the case of HD\,75049 \citep{Kochukhovetal2015}, on account of the low
inclination $i=30\degr$, it will be much more difficult to obtain the direct
proof of a violation of the force-free condition despite the extreme field
strengths involved. It is however already amply clear that the Tikhonov
regularization functional applied in the analysis of HD\,32633 and of
HD\,119419 does not lead to solutions that are physically -- instead of merely
mathematically -- feasible. It comes as a serious blow to the credibility of
ZDM that for CP stars with strong fields, it routinely converges to physically
impossible magnetic geometries and corresponding completely spurious abundance
maps, while displaying good or excellent fits to the observed Stokes IQUV
profiles. No doubt, we are faced with particularly obnoxious instances of
non-unique inversion results.

\section{Ever changing maps}
\label{mapchange}

Regularization and force-free fields are not the only problem facing ZDM.
The permanent changes of magnetic and abundance maps pervading the
literature are equally worrisome. Let us start with 53\,Cam which has
been observed in all 4 Stokes parameters just once, but analyzed and
reanalyzed at least 5 times with widely different results for the magnetic
field modulus (Table\,\ref{53Cam}). Contrasts range from 21.9 to 29.5\,kG,
minimum field strengths from 1.4 to 4.0\,kG, and maximum field strengths
from 25.1 to 32.3\,kG; At the same time, still based on the same data set,
abundance contrasts for Si and Fe change by about 1\,dex;
$\Delta\epsilon$(Si)$=3.4$\,dex and $\Delta\epsilon$(Fe)$=4.0$\,dex
\citep{KochukhovKoBaWaetal2004} become $\Delta\epsilon$(Si)$=4.3$\,dex
and $\Delta\epsilon$(Fe)$=5.0$\,dex \citep{Piskunov2008}. All
these changes are far outside the error limits of ZDM claimed by
\citet{Kochukhov2017}.

53\,Cam is not an isolated case. $\alpha^2$CVn has been observed and analyzed
a number of times and there are at least 9 different magnetic geometries to
be found in the literature (Table\,\ref{alpha2}). The field modulus contrasts
range between 2.5 and 6.3\,kG. the minimum field strengths between 0.0 and
1.4\,kG, and the maximum field strengths between 3.1 and 7.7\,kG, a factor
of 2.5\,! Earlier magnetic mappings based on Stokes $IV$ had determined dipolar
plus quadrupolar contributions of $B_{\rm d} = 5.3$ and $B_{\rm q} = 1.5$\,kG
\citep{KochPisketal2001}, soon to be replaced by $B_{\rm d} = 6.18$ and
$B_{\rm q} = 1.08$\,kG \citep{KochukhovKoPiIlTu2002}. These impressive
discrepancies in the magnetic field are accompanied by substantial changes
in the recovered abundance contrasts: from 2.4\,dex to 4.0\,dex (Si),
from 4.8 to 3.0\,dex (Cl), from 1.8\,dex to 4.0\,dex (Ti), from 2.3 to
3.5\,dex (Fe), and from 2.0 to 4.0\,dex (Nd). Since the abundances have
always been derived simultaneously with the magnetic field, it would seem
that only the neglected local atmospheric structure can be considered
responsible if one followed the arguments of \citet{Kochukhov2017}. 
Abundance changes of 2\,dex and more however are in complete contradiction
to his claims that ignoring the varying local atmospheres leads to average
reconstruction errors of $\approx 0.2$\,dex and maximum errors of
$\approx 0.3$\,dex and that the {\sc invers} code achieves an overall
average accuracy of 0.06–0.09\,dex and maximum errors of 0.12\,dex. The
conclusion is inevitable that there must be other sources of errors than
those discussed by \citet{Kochukhov2017}.

\begin{deluxetable}{@{}ccl}\label{53Cam}
\tablecaption{Magnetic field moduli for 53\,Cam.}
\tablewidth{0pt}
\tablehead{
\colhead{$B_{\rm min}$\,[kG]} & \colhead{$B_{\rm max}$\,[kG]}& \colhead{Reference} 
}
\startdata
2.8 & 32.3 & \citep{PiskunovKoch2003}\\
3.8 & 28.2 & \citep{Kochukhovetal2003}\\
4.0 & 25.9 & \citep{KochukhovKoBaWaetal2004}\\
1.4 & 26.1 & \citep{Kochukhov2004IAUS224}\\
1.5 & 25.1 & \citep{KochukhovPhysicsMagnStars2007}\\
\enddata
\tablecomments{Moduli determined from the same set of Stokes $IQUV$ observations.}
\end{deluxetable}

May we mention at this point that for another frequently observed and
analyzed star, HD\,24712, abundances are beset by similar problems.
\citet{LuftingerLuKoRyetal2010b} have claimed a Nd contrast of 1.1\,dex;
although in a later study it is stated that “the new maps confirm the
previous findings, and also show some details the previous study lacked''
\citep{RusomarovRuKoRyetal2015}, it emerges that not only does the Nd
contrast increase to 2.6\,dex, but the spot even changes position from about
$-25\degr$ south to near $-75\degr$ south! Finally, it is worthwhile to have
a look at HR\,3831 where the same set of observations has led to sometimes
widely different abundance maps -- keep in mind that the magnetic field
was not determined, but simply assumed dipolar and subsequently disregarded
in the Doppler mapping procedure. Putting aside the insoluble problem of
the structure of a hypothetical atmosphere in a spot with oxygen more
abundant than hydrogen \citep{KochukhovKoDrRe2003} and its equilibrium
with the rest of the atmosphere where oxygen can be some 7\,dex\,(!!) less
abundant, we note a 2.0\,dex difference in the maximum abundance of Ca
between \citet{KochukhovKoDrRe2003} and \citet{KochukhovKoDrPiRe2004},
of 1.3\,dex for both Li and Eu, and an almost complete dissimilarity
between the respective Si maps.

\begin{figure*}
\begin{centering}
\includegraphics[width=102mm, height=88mm, angle=270]{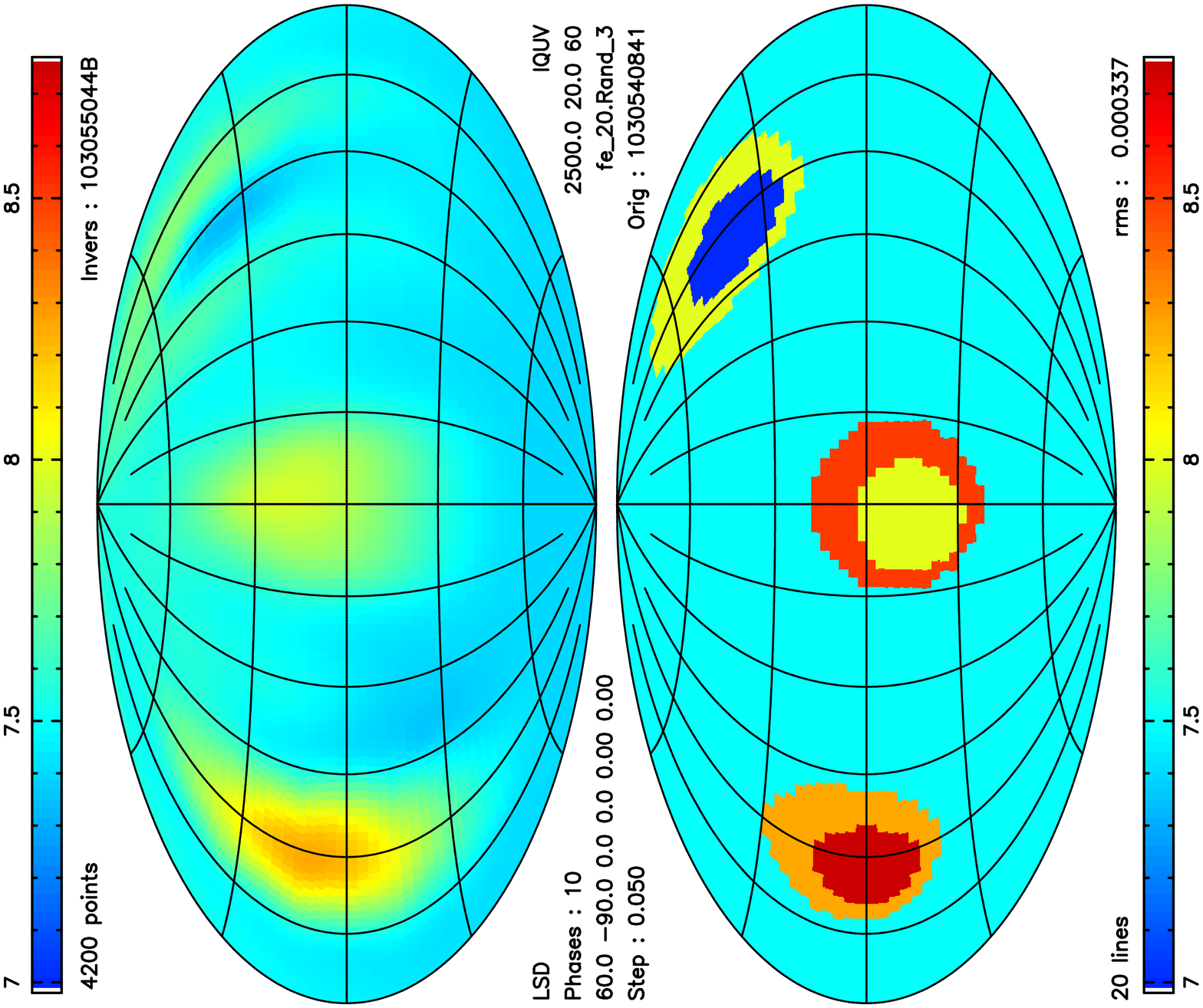}
\includegraphics[width=102mm, height=88mm, angle=270]{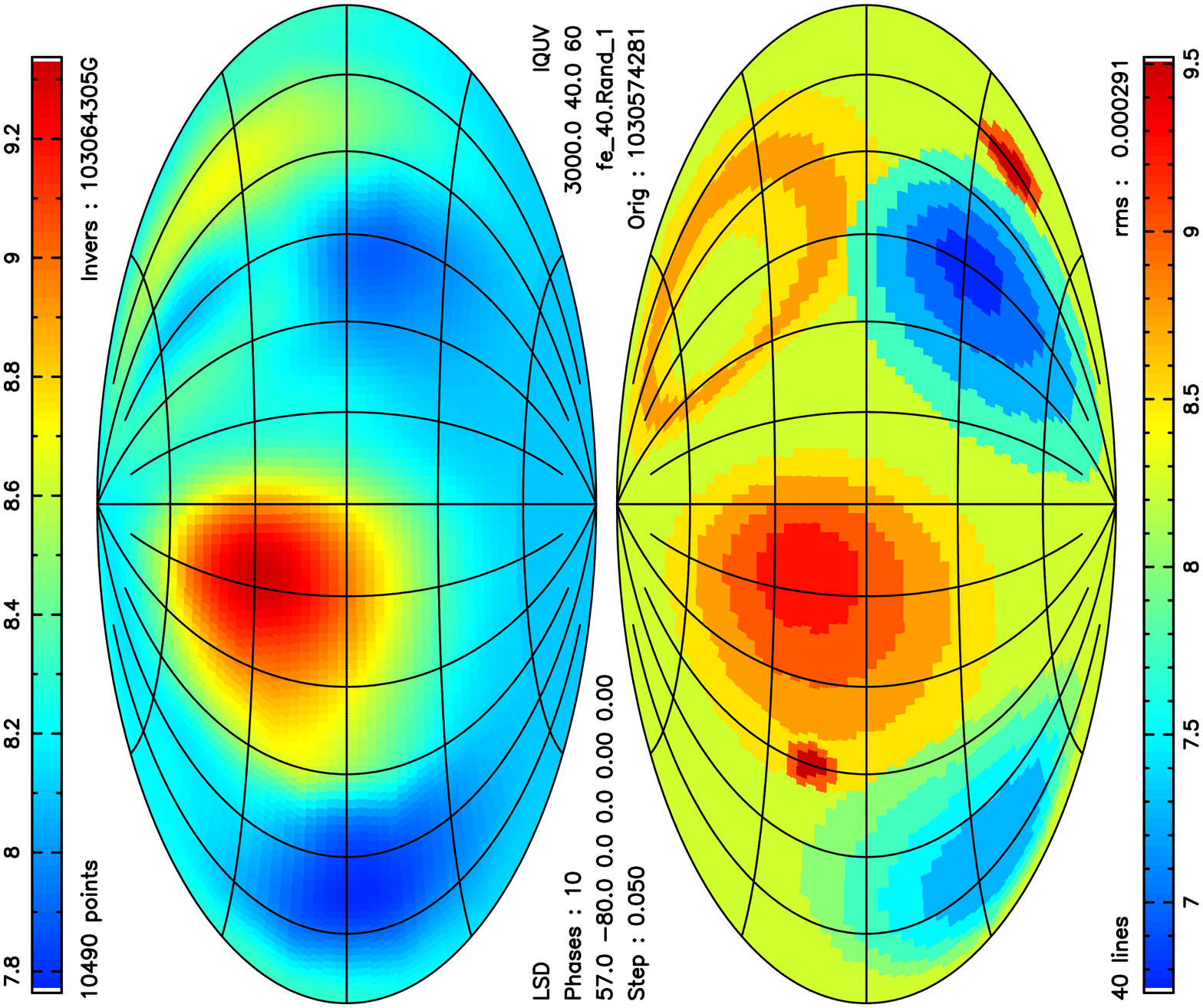}
\includegraphics[width=87mm,  height=41mm, angle=0, origin=l]{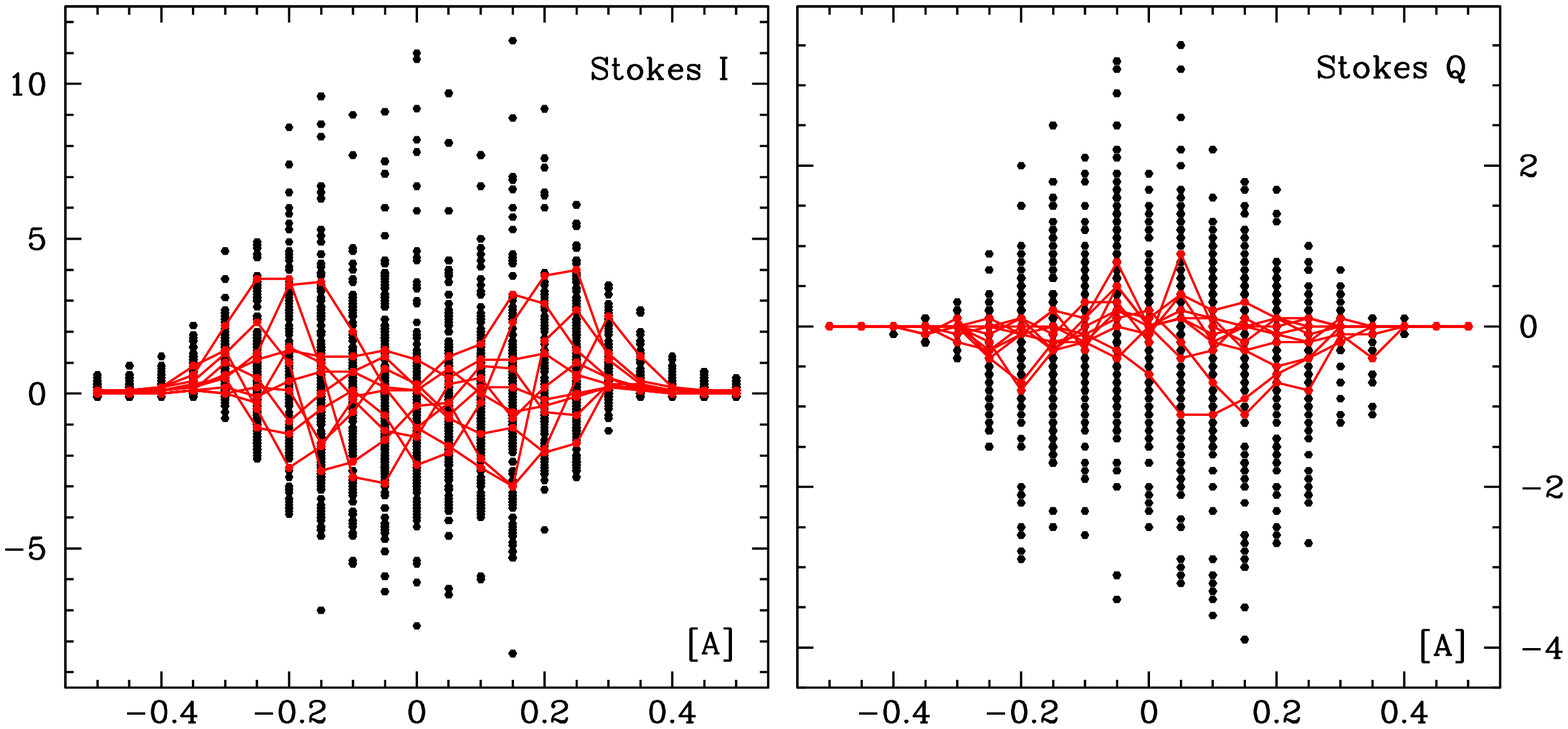}
\includegraphics[width=87mm,  height=41mm, angle=0, origin=r]{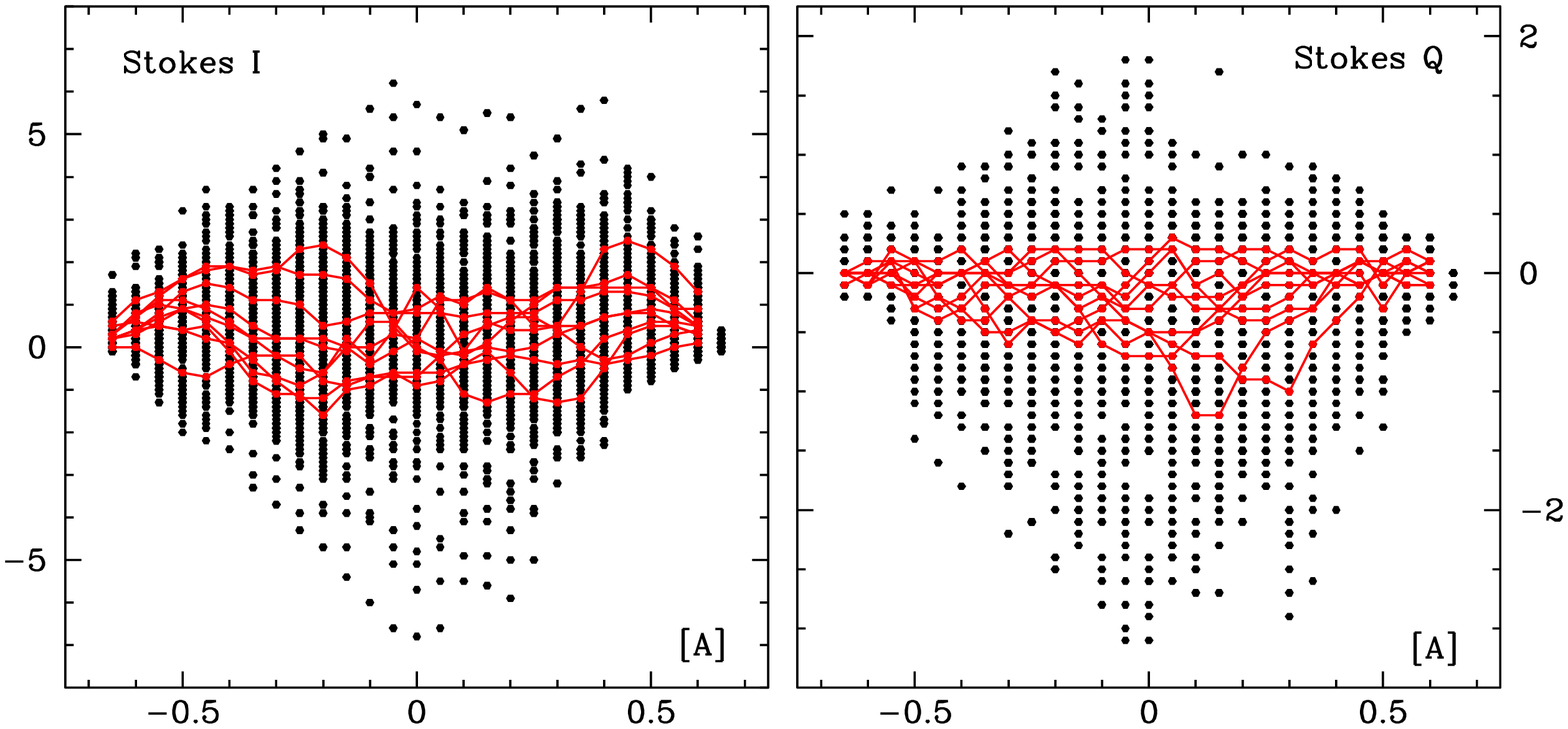}
\caption{
  LSD based ZDM results for 2 different assumed horizontal abundance distributions
  with a stellar atmosphere characterized by $T_{\rm eff} = 9000$\,K, $\log g = 4.0$
  and vertically constant chemical abundances.
  {\bf a)} Original and inverted abundance map based on 20 lines (left, middle and top)
  for a dipolar field with obliquity $90\degr$ and polar strength $B_{\rm p} = 5$\,kG.
  Inclination $i = 60\degr$, $v\,\sin i = 20\,{\rm km\,s}^{-1}$, wavelength resolution
  50\,m${\AA}$, 4500 pixel spatial grid. Bottom: residuals calculated minus ``observed''
  (in units of $10^{-3}$) for individual line profiles (black dots) and mean profiles
  (red line). {\bf b)} Original and inverted abundance map based on 40 lines (right,
  middle and top) for a dipolar field with obliquity $80\degr$, $B_{\rm p} = 6$\,kG,
  $i = 57\degr$, $v\,\sin i = 40\,{\rm km\,s}^{-1}$. Same spectral resolution and same
  spatial grid. Bottom : same residuals as before.
  }
\label{fig_LSD}
\end{centering}
\end{figure*}

\begin{deluxetable}{@{}cccl}\label{alpha2}
\tablecaption{Magnetic field moduli for $\alpha^2$CVn.}
\tablewidth{0pt}
\tablehead{
\colhead{$B_{\rm min}$\,[kG]} & \colhead{$B_{\rm max}$\,[kG]}& \colhead{Spectra}& \colhead{Reference} 
}
\startdata
1.4 & 7.7 &  $IV$  & \citep{Kochukhovetal2003}\\
1.0 & 4.3 & $IQUV$ & \citep{Kochukhov2004IAUS224}\\
0.6 & 3.1 & $IQUV$ & \citep{KochukhovPhysicsMagnStars2007}\\
0.0 & 4.4 & $IQUV$ & \citep{KochukhovKoWa2010}\\
0.1 & 4.9 & $IQUV$ & \citep{SilvesterSiKoWa2014a}\\
0.0 & 5.5 & $IQUV$ & \citep{SilvesterSiKoWa2014b}\\
0.5 & 4.5 & $IQUV$ & \citep{Kochukhov2018}\\
\enddata
\end{deluxetable}

The apex of unexplained published abundance differences is reached for
$\kappa$\,Psc. \citet{Ryabchikovaetal1996} finds Cr abundances between
-6.08 and -3.42 whereas according to \citet{PiskunovPietal1998} these
values become
-6.09 and +0.27, making Cr almost twice as abundant than hydrogen! The
authors do not agree either on inclination or gravity, $i = 35\degr$
and $\log g = 3.75$ being the choice of \citet{Ryabchikovaetal1996},
\citet{PiskunovPietal1998} adopting $i = 70\degr$ and $\log g = 4.50$.
The absurd consequence of a Cr abundance twice that of hydrogen would be
a pressure scale height in the spot that is about 30 times smaller than
in the rest of the atmosphere. Translated to our planet, the pressure on
top of Heaval (Barra, Outer Hebrides, 383m) would lie below the pressure
on top of Mt. Everest (8848m). How could such a huge pressure difference
remain stable for days, weeks, years or centuries, even when separated
by whole continents?

\section{LSD}
\label{LSD}

Least-squares deconvolution (LSD) \citep{Donatietal1997} was devised for
the detection of weak magnetic signals in noisy spectra. Its usefulness
for this particular purpose remains undisputed, although extreme care has
to be taken for the conversion of a LSD signal to quantities such as for
example $B_{\rm eff}$, the integrated longitudinal field of a magnetic star
\citep[see][]{Scaliaetal2017, Ramirez2020}. A lack of extended, realistic
tests makes it as yet impossible to assess whether the use of LSD mean
profiles in ZDM as carried out by \citet{KochukhovKoLuNeAl2014} leads to
valid stellar maps. Please keep in mind that LSD based ZDM is just one
special instance of single-line inversions which only for highly idealized
cases have been shown to result in unique solutions. \citet{StiftLeone2017a}
for example have demonstrated that one is frequently/usually faced with
multiple solutions even in the total absence of photon noise and despite
almost perfect fits to the observed profiles. The use of LSD profiles in ZDM
of course drastically lowers the observational noise, but unfortunately this
is done at the expense of the information contained in the various spectral
lines (different strengths, Zeeman patterns, ...).

To see what can happen, let us have a look at a star featuring 2 medium-size
spots featuring each 2 abundances, and 1 ring-like structure. Adopting an
inclination of $i = 60\degr$, a centered dipole normal to the rotational
axis with polar strength 5.0\,kG and a projected rotational velocity of
$v\,\sin i = 20\,{\rm km\,s}^{-1}$, we synthesize spectra for 20 unblended
lines at 10 equidistant phases at a wavelength resolution of 0.050\,{\AA}
and spatial resolution of $\approx 4500$ surface pixels. The subsequent
inversion -- based on all 4 Stokes $IQUV$ parameters -- is carried out under
the assumptions that the stellar parameters like temperature, gravity,
magnetic field etc., but also the atomic parameters, are exactly known.
The only unknowns are the horizontal variations of the chemical abundance
(taken vertically constant) of a single element. There are thus some 4000
unknowns to be derived from the phase-dependent $IQUV$ profiles of a single
mean line; for each of the 10 phases, the mean line results from 4100
individual contributions in $IQUV$. Despite the large number (16800) of
profile points, thanks to LSD we are faced with a classical, heavily
underdetermined inverse problem which has to be solved based on a mere 840
mean profile points. We follow the approach of \citet{KochukhovKoLuNeAl2014}
and make the fit to the observed LSD profiles with the help of mean profiles
determined at each iteration step from the individually synthesized spectral
lines. It is thus not assumed that the LSD profiles display the behavior of
a hypothetical single spectral line with some ill-defined mean parameters.
Although the fit to the observed LSD $IQUV$ profiles is good to a few $10^{-4}$,
Figure\,\ref{fig_LSD}a (top and middle) reveals a disappointing map that
indicates the presence of 2 low-contrast spots, giving also a marginal hint
at a possible ring-like structure, but failing outright to yield correct
abundances. The minimum abundance is overestimated by 0.33\,dex, the maximum
abundance underestimated by 0.50\,dex so that the original 1.75\,dex contrast
is almost halved to 0.92\,dex\,!

In another test, we worked with a star featuring 5 structured spots of varying
size and 1 large ring-like structure. Adopting inclination and obliquity of
the centered dipole similar to those of the former model, we increased the
polar magnetic field strength to 6.0\,kG, the projected rotational velocity
to $v\,\sin i = 40\,{\rm km\,s}^{-1}$ and the number of spectral lines to 40.
LSD reduces the resulting $42\,000$ profile points to 1040, leading to an
abundance map which essentially recovers only the strongest spot at $+19\degr$
despite a fit that is again good to a few $10^{-4}$ (Figure\,\ref{fig_LSD}b, top
and middle). All three southern spots would remain invisible unless one uses
different color scales for original and ZDM map respectively as in this plot.
Thanks to the chosen color scales one can see that the abundances in the 2 large
southern spots are overestimated by up to 1\,dex, and that the small overabundant
spots at $-40\degr$ and $+17\degr$ do not show up at all. In addition, we note
that the underabundant southern spots shift from $-32\degr$ to $\approx -14\degr$
and from $-33\degr$ to the equator.

\begin{figure}
\begin{centering}
\includegraphics[width=62mm, height=40mm, angle=0]{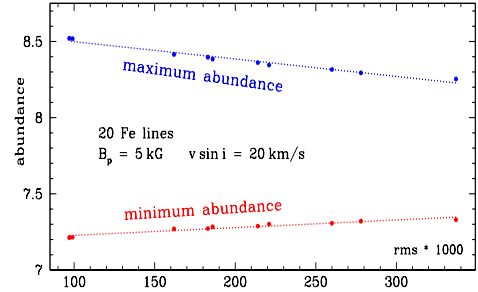}
\caption{
  Recovered abundance contrast as a function of the rms scatter of the fit
  to the 4 Stokes LSD profiles.
  }
\label{fig_5044}
\end{centering}
\end{figure}

A detailed analysis of the mean ZDM profiles and the profiles of the
individual lines reveals the disturbing fact that some kind of numerical
compensation mechanism can ensure almost perfect fits between input and
modeled mean LSD Stokes profiles, although the fit to the individual spectral
lines may be quite bad. This is demonstrated in Figure\,\ref{fig_LSD}a (bottom)
where we have plotted the differences calculated minus ``observed'' (input) for
the individual line profiles (black dots) and for the mean profile (red line),
both in Stokes $I$ and $Q$. It has to be emphasized that these differences,
exceeding 1\% of the continuum in Stokes $I$, do not represent observational
errors but are entirely due to incorrect ZDM abundance maps. As we can see, it
happens that individual lines are affected in different ways and that errors
thus compensate, leading to excellent mean fits as in the Stokes $Q$ example.
Our second model displays quite similar behavior, Stokes $Q$ being particularly
instructive (Figure\,\ref{fig_LSD}b, bottom).

Is there any information left in the mean Stokes profiles which would allow
us to correctly recover the spots in Figure\,\ref{fig_LSD}a and their true
abundances? We think that we can confidently state that this information has
not been destroyed in the LSD procedure, but that in real life it remains
inaccessible. Figure\,\ref{fig_5044} shows how in the course of the ZDM
iterations the inverted minimum and maximum abundances change with the quality
of the fit. At a rms scatter of the fit of $3.37\times10^{-4}$, the 2 spots
are only poorly recovered and the ring-like structure hardly recognizable.
Not before the rms scatter has come down to $9.7\times10^{-5}$ does the LSD-ZDM
inversion start to yield an abundance map that resembles the input map, but
the contrast of 1.3\,dex is still far from the original 1.75\,dex. In other
words, even fitting observed Stokes profiles to an accuracy of $10^{-3}$ or a
few $10^{-4}$ will not suffice to glean information on this simple surface
structure.

From these results it transpires that LSD based ZDM makes the traditional
problem of single-line inversions even less tractable: the code now can
achieve an almost perfect fit to the mean observed profiles by summing up
over individual profiles which may largely be incorrect. One can encounter
at least 3 major situations:\\
1) The fit to the LSD profile is excellent, but differences between the
   synthetic and the observed individual spectral lines are much larger.
   The resulting map is in error because LSD has introduced many additional
   degrees of freedom so that it becomes more likely that ZDM converges to
   spurious and unphysical solutions (as in Figure\,\ref{fig_LSD}a).\\
2) Both the fits to the individual spectral lines and to the LSD profiles
   are excellent but the resulting map is entirely at variance with the
   input map (as in Figure\,\ref{fig_LSD}b).\\
3) All fits are excellent and the final map is correct.\\
Unfortunately, case 2) implies that even an almost perfect fit to all 40
lines individually can result in a spurious map. We do not at present see
how it would be possible to discern between 2) and 3) in a LSD based ZDM
inversion.

\section{Conclusions}
\label{conclusions}

In the previous sections we have looked closely at the regularization
employed by the {\sc invers} family of codes for ZDM of CP stars, both
regarding the vector magnetic field and the abundance inhomogeneities.
Whether chemical profiles are assumed variable in depth but horizontally
homogeneous, or taken to correspond to step-like functions that can
change in all three dimensions, the verdict is clear: the regularization
functionals proposed by Kochukhov and collaborators do not in the least
reflect anything we know about the physics of the atmospheres of magnetic
CP stars or of our Sun.

In the past, the magnetic geometries as derived from Stokes $IV$ observations
or later from high resolution and high signal-to-noise ratio Stokes $IQUV$
line profiles have escaped close scrutiny. Although the published maps for
53\,Cam  did not fulfill the condition of zero divergence, there never was
a reassessment based on new observations. Later analyses of strongly magnetic
BpAp stars took more care to ensure zero divergence by fitting spherical
harmonics to the magnetic vector field, but as we have shown, the formulae
used are incompatible with force-free fields. Looking at the huge differences
between the published horizontal field maps and the horizontal fields derived
from the radial maps, there can be little doubt that all these published
abundance maps have to be discarded.

The ever-changing magnetic and abundance maps for several well observed and
well analyzed magnetic CP stars are not apt either to increase our confidence
in the alleged accuracy of ZDM inversions.There are also those alleged
low-contrast Cu and Ni spots in HD\,50773 \citep{LuftingerLuFrWeetal2010a}
that can be shown to exhibit spectral signatures completely swamped by the
noise of the observed spectra \citep{StiftLeone2017b}. On the other side,
when O, Si, or Cr are claimed to be more abundant than hydrogen, and other
metals like Mn and Fe almost as abundant as helium, how could these spots
ever stay in equilibrium with the surrounding atmosphere? How could such
extreme local atmospheres be treated correctly both physically and numerically?

In view of the results obtained in the course of this study, the bleak outlook
for ZDM presented by \citet{StiftLeone2017a} appears to have become hopeless.
The problem of a regularization functional that would reflect the physics of
stratified magnetic CP star atmospheres admits of no solution. We are faced
with a 3D problem of horizontal pressure and vertical hydrostatic equilibrium
that cannot be adequately approximated by a 1D or 2D approach involving isolated
“cylinders” characterized by some magnetic field vector. In the course of time,
abundance gradients build up in each ``cylinder'' -- depending on the local
field strength and angle -- density, pressure and temperature of the local
atmosphere change differently in each ``cylinder'', but what happens with the
ensuing horizontal differences in gas pressure? Equilibrium must be reached
on the shortest of time scales between one ``cylinder'' and {\em all} its
neighbors, not just the nearest. A strong vertical field may help to stabilize
the ``cylinders'', but nearly horizontal field lines will not do so. It is
inconceivable, even with the most sophisticated codes and the most powerful
supercomputers, to simulate the field-dependent build-up in time of hundreds
or thousands of vertical abundance profiles, and their global, 3D horizontal
interaction during or after each time step, once again as a function of the
local magnetic field vector. It is even less conceivable to do this for every
possible magnetic geometry. Only entirely new techniques will enable us one day
to reliably unravel the mystery of CP star abundance and magnetic maps.

\section*{Acknowledgments}

MJS wants to express his gratitude to Dr.~S.~Bagnulo for having reawakened
his interest in ZDM and for having initiated a collaboration involving
Armagh Observatory. Thanks go in particular to its former director, Prof.
M.~Bailey for his unflinching support, including the extensive use of
dedicated servers for diffusion calculations. We are most grateful to
Dr.~Patrick Alken (CIRES, Boulder, CO) and to Dr.~Ciaran D. Beggan (British
Geological Survey, Edinburgh) for clarifying issues concerning spherical
harmonic magnetic models of the earth and their application to magnetic CP
stars. Dr.~Dennis Westra (https://www.mat.univie.ac.at/\~westra/) patiently
introduced MJS to some of the intricacies of the associated Legendre
functions. We acknowledge financial contribution from the {\it Programma
ricerca di Ateneo UNICT 2020-22 linea 2}. This work would never have been
possible without the marvelous public GNAT Ada compiler of AdaCore.

\bibliographystyle{aasjournal.bst}
\bibliography{ZDM}

\begin{thebibliography}{}
\expandafter\ifx\csname natexlab\endcsname\relax\def\natexlab#1{#1}\fi
\providecommand{\url}[1]{\href{#1}{#1}}
\providecommand{\dodoi}[1]{doi:~\href{http://doi.org/#1}{\nolinkurl{#1}}}
\providecommand{\doeprint}[1]{\href{http://ascl.net/#1}{\nolinkurl{http://ascl.net/#1}}}
\providecommand{\doarXiv}[1]{\href{https://arxiv.org/abs/#1}{\nolinkurl{https://arxiv.org/abs/#1}}}

\bibitem[{{Alecian}(2015)}]{Alecian2015}
{Alecian}, G. 2015, MNRAS, 454, 3143, \dodoi{10.1093/mnras/stv2205}

\bibitem[{{Alecian} \& {Stift}(2002)}]{AlecianAlSt2002}
{Alecian}, G., \& {Stift}, M.~J. 2002, \aap, 387, 271,
  \dodoi{10.1051/0004-6361:20020381}

\bibitem[{{Alecian} \& {Stift}(2004)}]{AlecianAlSt2004}
---. 2004, A\&A, 416, 703, \dodoi{10.1051/0004-6361:20034457}

\bibitem[{{Alecian} \& {Stift}(2006)}]{AlecianAlSt2006i}
---. 2006, A\&A, 454, 571, \dodoi{10.1051/0004-6361:20054558}

\bibitem[{{Alecian} \& {Stift}(2007)}]{AlecianAlSt2007g}
---. 2007, A\&A, 475, 659, \dodoi{10.1051/0004-6361:20078000}

\bibitem[{{Alecian} \& {Stift}(2010)}]{AlecianAlSt2010}
---. 2010, A\&A, 516, A53+, \dodoi{10.1051/0004-6361/200913772}

\bibitem[{{Alecian} \& {Stift}(2017)}]{AlecianStift2017}
---. 2017, \mnras, 468, 1023, \dodoi{10.1093/mnras/stx496}

\bibitem[{{Alecian} \& {Stift}(2019)}]{AlecianStift2019}
---. 2019, \mnras, 482, 4519, \dodoi{10.1093/mnras/sty3003}

\bibitem[{{Alecian} {et~al.}(2011){Alecian}, {Stift}, \&
  {Dorfi}}]{AlecianAlStDo2011}
{Alecian}, G., {Stift}, M.~J., \& {Dorfi}, E.~A. 2011, MNRAS, 418, 986

\bibitem[{{Alecian} \& {Vauclair}(1981)}]{AlecianAlVa1981}
{Alecian}, G., \& {Vauclair}, S. 1981, A\&A, 101, 16

\bibitem[{{Alken} {et~al.}(2021){Alken}, {Th{\'e}bault}, {Beggan}, {Amit},
  {Aubert}, {Baerenzung}, {Bondar}, \& {Brown}}]{IGRF13}
{Alken}, P., {Th{\'e}bault}, E., {Beggan}, C.~D., {et~al.} 2021, Earth,
  Planets, and Space, 73, 49, \dodoi{10.1186/s40623-020-01288-x}

\bibitem[{{Babel} \& {Michaud}(1991{\natexlab{a}})}]{BabelMichaud1991a}
{Babel}, J., \& {Michaud}, G. 1991{\natexlab{a}}, \apj, 366, 560,
  \dodoi{10.1086/169591}

\bibitem[{{Babel} \& {Michaud}(1991{\natexlab{b}})}]{BabelMichaud1991b}
---. 1991{\natexlab{b}}, \aap, 241, 493

\bibitem[{{Donati} {et~al.}(1997){Donati}, {Semel}, {Carter}, {Rees}, \&
  {Collier Cameron}}]{Donatietal1997}
{Donati}, J.-F., {Semel}, M., {Carter}, B.~D., {Rees}, D.~E., \& {Collier
  Cameron}, A. 1997, MNRAS, 291, 658

\bibitem[{{Kochukhov}(2004)}]{Kochukhov2004IAUS224}
{Kochukhov}, O. 2004, in The A-Star Puzzle, ed. J.~{Zverko}, J.~{Ziznovsky},
  S.~J. {Adelman}, \& W.~W. {Weiss}, Vol. 224, 433,
  \dodoi{10.1017/S1743921304004855}

\bibitem[{{Kochukhov}(2007)}]{KochukhovPhysicsMagnStars2007}
{Kochukhov}, O. 2007, in Physics of Magnetic Stars, ed. I.~I. {Romanyuk}, D.~O.
  {Kudryavtsev}, O.~M. {Neizvestnaya}, \& V.~M. {Shapoval}, 61

\bibitem[{{Kochukhov}(2017)}]{Kochukhov2017}
{Kochukhov}, O. 2017, \aap, 597, A58, \dodoi{10.1051/0004-6361/201629768}

\bibitem[{Kochukhov(2018)}]{Kochukhov2018}
Kochukhov, O. 2018, Stellar Magnetic Fields, ed. J.~Sánchez~Almeida \& M.~J.
  Martínez~González, Canary Islands Winter School of Astrophysics (Cambridge
  University Press), 47

\bibitem[{{Kochukhov} {et~al.}(2004{\natexlab{a}}){Kochukhov}, {Bagnulo},
  {Wade}, {Sangalli}, {Piskunov}, {Landstreet}, {Petit}, \&
  {Sigut}}]{KochukhovKoBaWaetal2004}
{Kochukhov}, O., {Bagnulo}, S., {Wade}, G.~A., {et~al.} 2004{\natexlab{a}},
  A\&A, 414, 613, \dodoi{10.1051/0004-6361:20031595}

\bibitem[{{Kochukhov} {et~al.}(2003{\natexlab{a}}){Kochukhov}, {Drake}, \& {de
  La Reza}}]{KochukhovKoDrRe2003}
{Kochukhov}, O., {Drake}, N.~A., \& {de La Reza}, R. 2003{\natexlab{a}}, in
  Modelling of Stellar Atmospheres, ed. N.~{Piskunov}, W.~W. {Weiss}, \& D.~F.
  {Gray}, Vol. 210, D22

\bibitem[{{Kochukhov} {et~al.}(2004{\natexlab{b}}){Kochukhov}, {Drake},
  {Piskunov}, \& {de la Reza}}]{KochukhovKoDrPiRe2004}
{Kochukhov}, O., {Drake}, N.~A., {Piskunov}, N., \& {de la Reza}, R.
  2004{\natexlab{b}}, A\&A, 424, 935, \dodoi{10.1051/0004-6361:20040517}

\bibitem[{{Kochukhov} {et~al.}(2014){Kochukhov}, {L{\"u}ftinger}, {Neiner},
  {Alecian}, \& {MiMeS Collaboration}}]{KochukhovKoLuNeAl2014}
{Kochukhov}, O., {L{\"u}ftinger}, T., {Neiner}, C., {Alecian}, E., \& {MiMeS
  Collaboration}. 2014, \aap, 565, A83, \dodoi{10.1051/0004-6361/201423472}

\bibitem[{{Kochukhov} \& {Piskunov}(2002)}]{KochukhovKoPi2002}
{Kochukhov}, O., \& {Piskunov}, N. 2002, A\&A, 388, 868,
  \dodoi{10.1051/0004-6361:20020300}

\bibitem[{{Kochukhov} {et~al.}(2003{\natexlab{b}}){Kochukhov}, {Piskunov},
  {Bagnulo}, {Landstreet}, {Sigut}, {Petit}, \& {Wade}}]{Kochukhovetal2003}
{Kochukhov}, O., {Piskunov}, N., {Bagnulo}, S., {et~al.} 2003{\natexlab{b}}, in
  Astronomical Society of the Pacific Conference Series, Vol. 307, Solar
  Polarization, ed. J.~{Trujillo-Bueno} \& J.~{Sanchez Almeida}, 549

\bibitem[{{Kochukhov} {et~al.}(2001){Kochukhov}, {Piskunov}, {Ilyin}, {Ilyina},
  \& {Tuominen}}]{KochPisketal2001}
{Kochukhov}, O., {Piskunov}, N., {Ilyin}, I., {Ilyina}, S., \& {Tuominen}, I.
  2001, in Astronomical Society of the Pacific Conference Series, Vol. 248,
  Magnetic Fields Across the Hertzsprung-Russell Diagram, ed. G.~{Mathys},
  S.~K. {Solanki}, \& D.~T. {Wickramasinghe}, 321

\bibitem[{{Kochukhov} {et~al.}(2002){Kochukhov}, {Piskunov}, {Ilyin}, {Ilyina},
  \& {Tuominen}}]{KochukhovKoPiIlTu2002}
{Kochukhov}, O., {Piskunov}, N., {Ilyin}, I., {Ilyina}, S., \& {Tuominen}, I.
  2002, \aap, 389, 420, \dodoi{10.1051/0004-6361:20020299}

\bibitem[{{Kochukhov} \& {Ryabchikova}(2018)}]{KochRyab2018}
{Kochukhov}, O., \& {Ryabchikova}, T.~A. 2018, \mnras, 474, 2787,
  \dodoi{10.1093/mnras/stx2961}

\bibitem[{{Kochukhov} {et~al.}(2017){Kochukhov}, {Silvester}, {Bailey},
  {Landstreet}, \& {Wade}}]{KochukhovKoSiBaLaWa2017}
{Kochukhov}, O., {Silvester}, J., {Bailey}, J.~D., {Landstreet}, J.~D., \&
  {Wade}, G.~A. 2017, \aap, 605, A13, \dodoi{10.1051/0004-6361/201730919}

\bibitem[{{Kochukhov} {et~al.}(2006){Kochukhov}, {Tsymbal}, {Ryabchikova},
  {Makaganyk}, \& {Bagnulo}}]{KoTsRyMaBa2006}
{Kochukhov}, O., {Tsymbal}, V., {Ryabchikova}, T., {Makaganyk}, V., \&
  {Bagnulo}, S. 2006, A\&A, 460, 831, \dodoi{10.1051/0004-6361:20065607}

\bibitem[{{Kochukhov} \& {Wade}(2010)}]{KochukhovKoWa2010}
{Kochukhov}, O., \& {Wade}, G.~A. 2010, ApJ, 513, A13,
  \dodoi{10.1051/0004-6361/200913860}

\bibitem[{{Kochukhov} {et~al.}(2015){Kochukhov}, {Rusomarov}, {Valenti},
  {Stempels}, {Snik}, {Rodenhuis}, {Piskunov}, {Makaganiuk}, {Keller}, \&
  {Johns-Krull}}]{Kochukhovetal2015}
{Kochukhov}, O., {Rusomarov}, N., {Valenti}, J.~A., {et~al.} 2015, \aap, 574,
  A79, \dodoi{10.1051/0004-6361/201425065}

\bibitem[{{L{\"u}ftinger} {et~al.}(2010{\natexlab{a}}){L{\"u}ftinger},
  {Kochukhov}, {Ryabchikova}, {Piskunov}, {Weiss}, \&
  {Ilyin}}]{LuftingerLuKoRyetal2010b}
{L{\"u}ftinger}, T., {Kochukhov}, O., {Ryabchikova}, T., {et~al.}
  2010{\natexlab{a}}, A\&A, 509, A71, \dodoi{10.1051/0004-6361/200811545}

\bibitem[{{L{\"u}ftinger} {et~al.}(2010{\natexlab{b}}){L{\"u}ftinger},
  {Fr{\"o}hlich}, {Weiss}, {Petit}, {Auri{\`e}re}, {Nesvacil}, {Gruberbauer},
  {Shulyak}, {Alecian}, {Baglin}, {Baudin}, {Catala}, {Donati}, {Kochukhov},
  {Michel}, \& et. al.}]{LuftingerLuFrWeetal2010a}
{L{\"u}ftinger}, T., {Fr{\"o}hlich}, H.-E., {Weiss}, W.~W., {et~al.}
  2010{\natexlab{b}}, A\&A, 509, A43, \dodoi{10.1051/0004-6361/200912239}

\bibitem[{{Michaud} {et~al.}(1981){Michaud}, {Charland}, \&
  {Megessier}}]{MichaudMiChMe1981}
{Michaud}, G., {Charland}, Y., \& {Megessier}, C. 1981, A\&A, 103, 244

\bibitem[{{Nesvacil} {et~al.}(2012){Nesvacil}, {L{\"u}ftinger}, {Shulyak},
  {Obbrugger}, {Weiss}, {Drake}, {Hubrig}, {Ryabchikova}, {Kochukhov},
  {Piskunov}, \& {Polosukhina}}]{NesvacilNeLuShetal2012}
{Nesvacil}, N., {L{\"u}ftinger}, T., {Shulyak}, D., {et~al.} 2012, A\&A, 537,
  A151, \dodoi{10.1051/0004-6361/201117097}

\bibitem[{{Oksala} {et~al.}(2018){Oksala}, {Silvester}, {Kochukhov}, {Neiner},
  {Wade}, \& {MiMeS Collaboration}}]{Oksala_etal_2018}
{Oksala}, M.~E., {Silvester}, J., {Kochukhov}, O., {et~al.} 2018, \mnras, 473,
  3367, \dodoi{10.1093/mnras/stx2487}

\bibitem[{{Piskunov}(2001)}]{Piskunov2001}
{Piskunov}, N. 2001, in Astronomical Society of the Pacific Conference Series,
  Vol. 248, Magnetic Fields Across the Hertzsprung-Russell Diagram, ed.
  G.~{Mathys}, S.~K. {Solanki}, \& D.~T. {Wickramasinghe}, 293

\bibitem[{{Piskunov}(2008)}]{Piskunov2008}
{Piskunov}, N. 2008, Physica Scripta Volume T, 133, 014017,
  \dodoi{10.1088/0031-8949/2008/T133/014017}

\bibitem[{{Piskunov} \& {Kochukhov}(2002)}]{PiskunovPiKo2002}
{Piskunov}, N., \& {Kochukhov}, O. 2002, A\&A, 381, 736,
  \dodoi{10.1051/0004-6361:20011517}

\bibitem[{{Piskunov} {et~al.}(1998){Piskunov}, {Stempels}, {Ryabchikova},
  {Malanushenko}, \& {Savanov}}]{PiskunovPietal1998}
{Piskunov}, N., {Stempels}, H.~C., {Ryabchikova}, T.~A., {Malanushenko}, V., \&
  {Savanov}, I. 1998, Contributions of the Astronomical Observatory Skalnate
  Pleso, 27, 482

\bibitem[{{Piskunov} \& {Kochukhov}(2003)}]{PiskunovKoch2003}
{Piskunov}, N.~E., \& {Kochukhov}, O. 2003, in Astronomical Society of the
  Pacific Conference Series, Vol. 305, Magnetic Fields in O, B and A Stars:
  Origin and Connection to Pulsation, Rotation and Mass Loss, ed. L.~A.
  {Balona}, H.~F. {Henrichs}, \& R.~{Medupe}, 83

\bibitem[{{Ram{\'\i}rez V{\'e}lez}(2020)}]{Ramirez2020}
{Ram{\'\i}rez V{\'e}lez}, J.~C. 2020, \mnras, 493, 1130,
  \dodoi{10.1093/mnras/staa301}

\bibitem[{{Rusomarov}(2016)}]{Rusomarov2016}
{Rusomarov}, N. 2016, PhD thesis, Uppsala

\bibitem[{{Rusomarov} {et~al.}(2018){Rusomarov}, {Kochukhov}, \&
  {Lundin}}]{RusomarovRuKoLu2018}
{Rusomarov}, N., {Kochukhov}, O., \& {Lundin}, A. 2018, \aap, 609, A88,
  \dodoi{10.1051/0004-6361/201731914}

\bibitem[{{Rusomarov} {et~al.}(2016{\natexlab{a}}){Rusomarov}, {Kochukhov}, \&
  {Ryabchikova}}]{RusomarovRuKoRy2016}
{Rusomarov}, N., {Kochukhov}, O., \& {Ryabchikova}, T. 2016{\natexlab{a}},
  Astronomy and Astrophysics. Submitted

\bibitem[{{Rusomarov} {et~al.}(2016{\natexlab{b}}){Rusomarov}, {Kochukhov},
  {Ryabchikova}, \& {Ilyin}}]{RusomarovRuKoRyIl2016}
{Rusomarov}, N., {Kochukhov}, O., {Ryabchikova}, T., \& {Ilyin}, I.
  2016{\natexlab{b}}, \aap, 588, A138, \dodoi{10.1051/0004-6361/201527719}

\bibitem[{{Rusomarov} {et~al.}(2015){Rusomarov}, {Kochukhov}, {Ryabchikova}, \&
  {Piskunov}}]{RusomarovRuKoRyetal2015}
{Rusomarov}, N., {Kochukhov}, O., {Ryabchikova}, T., \& {Piskunov}, N. 2015,
  A\&A, 573, A123, \dodoi{10.1051/0004-6361/201424559}

\bibitem[{{Ryabchikova}(2014)}]{Ryabchikova2014}
{Ryabchikova}, T. 2014, in Putting A Stars into Context: Evolution,
  Environment, and Related Stars, ed. G.~{Mathys}, E.~R. {Griffin},
  O.~{Kochukhov}, R.~{Monier}, \& G.~M. {Wahlgren}, 220--228

\bibitem[{{Ryabchikova} {et~al.}(1996){Ryabchikova}, {Pavlova}, {Davydova}, \&
  {Piskunov}}]{Ryabchikovaetal1996}
{Ryabchikova}, T.~A., {Pavlova}, V.~M., {Davydova}, E.~S., \& {Piskunov}, N.~E.
  1996, Astronomy Letters, 22, 822

\bibitem[{{Scalia} {et~al.}(2017){Scalia}, {Leone}, {Gangi}, {Giarrusso}, \&
  {Stift}}]{Scaliaetal2017}
{Scalia}, C., {Leone}, F., {Gangi}, M., {Giarrusso}, M., \& {Stift}, M.~J.
  2017, \mnras, 472, 3554, \dodoi{10.1093/mnras/stx2090}

\bibitem[{{Silvester} {et~al.}(2017){Silvester}, {Kochukhov}, {Rusomarov}, \&
  {Wade}}]{SilvesterSiKoRuWa2017}
{Silvester}, J., {Kochukhov}, O., {Rusomarov}, N., \& {Wade}, G.~A. 2017,
  \mnras, 471, 962, \dodoi{10.1093/mnras/stx1606}

\bibitem[{{Silvester} {et~al.}(2014{\natexlab{a}}){Silvester}, {Kochukhov}, \&
  {Wade}}]{SilvesterSiKoWa2014}
{Silvester}, J., {Kochukhov}, O., \& {Wade}, G.~A. 2014{\natexlab{a}}, \mnras,
  440, 182, \dodoi{10.1093/mnras/stu306}

\bibitem[{{Silvester} {et~al.}(2014{\natexlab{b}}){Silvester}, {Kochukhov}, \&
  {Wade}}]{SilvesterSiKoWa2014a}
---. 2014{\natexlab{b}}, MNRAS, 440, 182, \dodoi{10.1093/mnras/stu306}

\bibitem[{{Silvester} {et~al.}(2014{\natexlab{c}}){Silvester}, {Kochukhov}, \&
  {Wade}}]{SilvesterSiKoWa2014b}
---. 2014{\natexlab{c}}, MNRAS, 444, 1442, \dodoi{10.1093/mnras/stu1531}

\bibitem[{{Silvester} {et~al.}(2015){Silvester}, {Kochukhov}, \&
  {Wade}}]{SilvesterSiKoWa2015}
---. 2015, \mnras, 453, 2163, \dodoi{10.1093/mnras/stv1775}

\bibitem[{{Stift}(1996)}]{Stift1996}
{Stift}, M.~J. 1996, in IAU Symposium, Vol. 176, Stellar Surface Structure, ed.
  K.~G. {Strassmeier} \& J.~L. {Linsky}, 61

\bibitem[{{Stift} \& {Alecian}(2009)}]{StiftStAl2009s}
{Stift}, M.~J., \& {Alecian}, G. 2009, MNRAS, 394, 1503,
  \dodoi{10.1111/j.1365-2966.2009.14419.x}

\bibitem[{{Stift} \& {Alecian}(2016)}]{StiftAlecian2016}
---. 2016, MNRAS, 457, 74, \dodoi{10.1093/mnras/stv2962}

\bibitem[{{Stift} {et~al.}(2013{\natexlab{a}}){Stift}, {Alecian}, \&
  {Dorfi}}]{StiftStAlDo2013}
{Stift}, M.~J., {Alecian}, G., \& {Dorfi}, E.~A. 2013{\natexlab{a}}, in EAS
  Publications Series, Vol.~63, EAS Publications Series, ed. G.~{Alecian},
  Y.~{Lebreton}, O.~{Richard}, \& G.~{Vauclair}, 227--232,
  \dodoi{10.1051/eas/1363026}

\bibitem[{{Stift} {et~al.}(2013{\natexlab{b}}){Stift}, {Hubrig}, {Leone}, \&
  {Mathys}}]{StiftStHuLeetal2013}
{Stift}, M.~J., {Hubrig}, S., {Leone}, F., \& {Mathys}, G. 2013{\natexlab{b}},
  in Astronomical Society of the Pacific Conference Series, Vol. 479, Progress
  in Physics of the Sun and Stars: A New Era in Helio- and Asteroseismology,
  ed. H.~{Shibahashi} \& A.~E. {Lynas-Gray}, 125

\bibitem[{{Stift} \& {Leone}(2017{\natexlab{a}})}]{StiftLeone2017a}
{Stift}, M.~J., \& {Leone}, F. 2017{\natexlab{a}}, \apj, 834, 24,
  \dodoi{10.3847/1538-4357/834/1/24}

\bibitem[{{Stift} \& {Leone}(2017{\natexlab{b}})}]{StiftLeone2017b}
---. 2017{\natexlab{b}}, \mnras, 465, 2880, \dodoi{10.1093/mnras/stw2885}

\bibitem[{{Stift} {et~al.}(2012){Stift}, {Leone}, \&
  {Cowley}}]{StiftStLeCo2012}
{Stift}, M.~J., {Leone}, F., \& {Cowley}, C.~R. 2012, MNRAS, 419, 2912,
  \dodoi{10.1111/j.1365-2966.2011.19933.x}

\bibitem[{{Tiwari}(2012)}]{Tiwari2012}
{Tiwari}, S.~K. 2012, \apj, 744, 65, \dodoi{10.1088/0004-637X/744/1/65}

\bibitem[{{Vauclair} {et~al.}(1979){Vauclair}, {Hardorp}, \&
  {Peterson}}]{VauclairVaHaPe1979}
{Vauclair}, S., {Hardorp}, J., \& {Peterson}, D.~M. 1979, ApJ, 227, 526,
  \dodoi{10.1086/156761}

\bibitem[{{Vogt} {et~al.}(1987){Vogt}, {Penrod}, \& {Hatzes}}]{VogtVoPeHa1987}
{Vogt}, S.~S., {Penrod}, G.~D., \& {Hatzes}, A.~P. 1987, ApJ, 321, 496,
  \dodoi{10.1086/165647}

\bibitem[{{Winch} {et~al.}(2005){Winch}, {Ivers}, {Turner}, \&
  {Stening}}]{WinchWiIvTuSt2005}
{Winch}, D.~E., {Ivers}, D.~J., {Turner}, J.~P.~R., \& {Stening}, R.~J. 2005,
  Geophysical Journal International, 160, 487,
  \dodoi{10.1111/j.1365-246X.2004.02472.x}

\end{thebibliography}

\end{document}